\PassOptionsToPackage{table,xcdraw}{xcolor}
\documentclass[manuscript,screen,acmlarge]{acmart}
\usepackage{multirow}
\usepackage{graphicx}
\usepackage{caption}
\usepackage{subcaption}
\usepackage{hyperref}
\usepackage{todonotes}

\newif\ifproofread
\newcommand{\vm}[1]{%
    \ifproofread
        \textcolor{blue}{#1}%
    \else
        #1%
    \fi
}
\proofreadfalse
\AtBeginDocument{%
  \providecommand\BibTeX{{%
    \normalfont B\kern-0.5em{\scshape i\kern-0.25em b}\kern-0.8em\TeX}}}

\setcopyright{acmcopyright}
\copyrightyear{2023}
\acmYear{2023}
\acmDOI{}

\acmJournal{JOCCH}

\begin{document}

\title[DoubleCheck]{DoubleCheck: Designing Community-based Assessability for Historical Person Identification}


\author{Vikram Mohanty}
\email{vikrammohanty@vt.edu}
\orcid{0000-0001-6296-3134} 
\affiliation{%
  \institution{Department of Computer Science, Virginia Tech}
  \city{Arlington}
  \state{Virginia}
  \country{USA}
  \postcode{22203}
}

\author{Kurt Luther}
\email{kluther@vt.edu}
\orcid{0000-0003-1809-6269} 
\affiliation{%
  \institution{Department of Computer Science, Virginia Tech}
  \city{Arlington}
  \state{Virginia}
  \country{USA}
  \postcode{22203}
}

\renewcommand{\shortauthors}{Mohanty, V. and Luther, K.}

\begin{abstract}
Historical photos are valuable for their cultural and economic significance, but can be difficult to identify accurately due to various challenges such as low-quality images, lack of corroborating evidence, and limited research resources. Misidentified photos can have significant negative consequences, including lost economic value, incorrect historical records, and the spread of misinformation that can lead to perpetuating conspiracy theories. To accurately assess the credibility of a photo identification (ID), it may be necessary to conduct investigative research, use domain knowledge, and consult experts. In this paper, we introduce DoubleCheck, a quality assessment framework for verifying historical photo IDs on Civil War Photo Sleuth (CWPS), a popular online platform for identifying American Civil War-era photos using facial recognition and crowdsourcing. \vm{DoubleCheck focuses on improving CWPS's user experience and system architecture to display information useful for assessing the quality of historical photo IDs on CWPS. In a mixed-methods evaluation of DoubleCheck, we found that users contributed a wide diversity of sources for photo IDs, which helped facilitate the community's assessment of these IDs through DoubleCheck's provenance visualizations. Further, DoubleCheck's quality assessment badges and visualizations supported users in making accurate assessments of photo IDs, even in cases involving ID conflicts.}
\end{abstract}



\maketitle

\section{Introduction}

Historical photo identification is important for verifying the identity of individuals depicted in the photos and providing context about the people, places, and events shown. Identifying people in historical photos helps in preserving cultural heritage, generates economic value~\cite{martinez_unknown_2012,schuessler_2017}, recognizes contributions of marginalized groups~\cite{Fortin_2018}, and helps connect people with their ancestors and cultural roots, which can be a powerful and meaningful experience~\cite{street_sears_smith_2021,roig-franzia_2018}. Accurately identifying and contextualizing historical photos helps us better understand and appreciate the events and people that have shaped our world~\cite{tucker_2020,tucker_2020_bonus,time2016time,tinkler2013using,scherer1990historical,hirsch2001surviving}.

However, the task of identifying historical photos can be a challenging problem due to a number of reasons that includes limited corroborating evidence (i.e., lacking context or accompanying information), aging and deterioration of the photos, limited resources for researching (i.e., domain knowledge, reference database, etc.), and potential inaccuracies that may have piled up over the years. There have been numerous instances of prominent historical photos getting misidentified and being corrected years later, such as the iconic World War II photographs of the US Marine Corps raising the flag at Iwo Jima~\cite{Schmidt_2018} and the sailor kissing a nurse in Times Square to celebrate the end of the war~\cite{slotnik_2019}. While the complexities of the task can contribute towards historical photos getting incorrectly identified~\cite{10.2307/27076640,10.2307/27172788}, there is also a possibility of intentional misidentification \vm{for financial incentives~\cite{handler2007retouching}}, to spawn conspiracy theories \cite{colimore_2019}, or sow disinformation in modern times \cite{evon_2020}. 

The proliferation of online platforms for history and genealogy discussion, such as Find-a-Grave, Ancestry.com, Facebook Groups, and Civil War Photo Sleuth, have made it easier for individuals with different areas of expertise to identify historical photos. Users of these platforms have been successful in archiving old records and rediscovering lost identifications~\cite{ruane_facebook_2014,mohanty2019photo}, but they are also susceptible to \emph{misidentification} due to limited expertise and over-reliance on imperfect automated tools like text search engines or face recognition~\cite{willever2014family,mohanty2020photo}. Therefore, it becomes important for the users of these online historical communities to be able to assess the authenticity of the crowdsourced information being shared on these platforms. However, verifying the identities of historical photos requires a combination of investigative research for corroborating multiple pieces of evidence, critical thinking, knowledge of the time period and context in which the photo was taken, and may also require consulting with other experts.

To address this problem, we introduce \emph{DoubleCheck}, a holistic quality assessment framework for supporting historical photo ID verifications. DoubleCheck builds on the \textit{assessable design} framework proposed by Forte et al.~\cite{forte2014designing}, based on the concepts of information provenance and stewardship. We focused on Civil War Photo Sleuth (CWPS)\footnote{\url{http://www.civilwarphotosleuth.com}}, an AI-infused online platform for identifying historical photos, which has over 17,000 registered users and over 25,000 identified Civil War portraits, and faces the problem of historical photo misidentification \cite{mohanty2019photo}. We first identified the challenges faced by the platform, and came up with specific goals to inform the design of DoubleCheck. We developed DoubleCheck on top of the existing CWPS platform to support its users in making accurate assessment of photo IDs. As part of DoubleCheck, we modified CWPS's architecture for capturing accurate provenance information, and incorporated domain knowledge to determine source trustworthiness, which was used for organizing the provenance information. Finally, we augmented the existing stewardship capabilities of CWPS by building an automated quality assessment engine as part of DoubleCheck, which combines community opinions and source trustworthiness to determine whether an ID is verified or not. 

We publicly released DoubleCheck on CWPS in the last quarter of 2020 and conducted a mixed-methods evaluation of twelve months of usage, which included interviews with potential users of different expertise levels, and log analysis of provenance and stewardship behaviors on the platform. We found that users were able to assess Civil War photo IDs better using the DoubleCheck framework. The stewardship visualizations for the quality assessment status of an ID boosted the confidence of users’ assessment. Users provided a wide range of provenance information, and found the organization of sources to be useful for their assessment.  

Our key contributions are:

\begin{itemize}
    \item a holistic quality assessment framework, \emph{DoubleCheck}, for supporting historical photo ID verification, 
    \item integrating DoubleCheck with Civil War Photo Sleuth (CWPS), an existing online history community, 
    \item a mixed-methods evaluation of \textit{ DoubleCheck} on CWPS after twelve months of deployment with real users.
\end{itemize}

We also discuss the implications of community participation, provenance and quality indicators on assessing historic photo IDs. 
\section{Related Work: Data Quality Assessment in Online Communities}

Over the years, numerous online communities, forums and websites have been developed for archiving and documenting history \cite{rosenzweig_can_2006}, generating family histories \cite{willever2014family,willever2012tell}, identifying and sharing historical photos \cite{mohanty2019photo}, trading antiques \cite{combs2005internet,heritage_auctions}, and facilitating discussions around history \cite{gilbert2020run}. Much like popular social media platforms such as Facebook and Twitter, these history-based platforms are also prone to the problem of misinformation, which may stem from multiple reasons including inexperienced contributors, platform's low bar of entry, and automation bias (i.e., over-reliance on automated suggestions) ~\cite{willever2014family,mohanty2020photo}. 

Rosenzweig \cite{rosenzweig_can_2006} analyzed Wikipedia as a source of historical scholarship, noting its policy against original research, and advocating for it as a tool for teaching the limitations of information sources and critical analysis of primary and secondary sources. Studies on journalism workflows have also thrown light on the importance of verifying source evidence for tackling misinformation~\cite{mcclure2022bridging,duffy2018naming,brandtzaeg2016emerging}, which is reflected in the design of several fact-checking tools and workflows that allow for easy visualization of the origin(s) of the information or the credibility of the source~\cite{zubiaga2018detection,diakopoulos2012finding,diakopoulos2021towards}.     

Motivated along similar lines, Forte et al \cite{forte2014designing} proposed the \textit{assessability} framework for designing assessable participatory information systems, based on information provenance and stewardship. The concept of provenance, extensively used in history and archival studies, describes information that makes it possible to trace the ownership or origins of the content, while stewardship refers to the processes that were used for maintaining the content, including its authorship. In the case of Wikipedia, Forte et al. found that visualizing provenance (i.e., citation types) and stewardship (i.e., article quality) had a significant impact on assessments of articles and Wikipedia as an information source. We draw on this work to incorporate the concepts of provenance (i.e., where did information about the photo ID come from?) and stewardship (i.e., how reliable is the photo ID information?) into the design of DoubleCheck. 

The Wikipedia visualizations were largely meant for establishing a proof-of-concept for \emph{assessable designs}, and therefore, showed example representations that were largely static (i.e., not created from live user data). Further, the stewardship visualizations essentially reflect an editor's assessment rating of the article quality, thus opening doors to subjective biases~\cite{yaari2011information}. As part of DoubleCheck, we built architectures to capture accurate provenance information from users and designed dynamic visualization to organize the source information according to their trustworthiness. DoubleCheck also mitigates pure subjective assessment by automatically combining source trustworthiness information with the community's opinions to determine whether an ID can be verified or not. 

Wiggins et al. \cite{wiggins2016community} investigated data validation practices on iNaturalist, an online platform dedicated to identifying photos of plant and animal species, and found the community's practices contributing towards information provenance through stewardship behaviors. While the log of user records are considered as provenance information, the stewardship aspects encompass user agreements on the organism's ID and the quality grade status of whether it’s a research grade identification or not (i.e., the results can be used for scientific research purposes), which the platform determines by two-third consensus by the community.

However, the task of identifying people in historical photos is fundamentally different from identifying plant and animal species in modern photos. For the latter, it might be possible to capture multiple high-resolution photos of multiple occurrences, which can further aid in the verification process. On the other hand, the same may not be true for historical photos; finding multiple photos of the same individual becomes more difficult as one goes back further in time and provenance information becomes highly critical for verifying the quality of a photo ID. Instead of placing sole importance on community consensus, DoubleCheck implements a nuanced rule-based policy that combines both community consensus and provenance information for verifying photo IDs. 

For maintaining the quality of its articles, Wikipedia relies on its editors following the Verifiability policy \cite{wikipedia_2021}, which states that all articles on Wikipedia must be verifiable, and all information on the page must be attributed to a reliable secondary source. Further, unsourced material should be flagged. While algorithmic solutions have been proposed to detect whether a statement within a Wikipedia article needs a citation source or not, ascertaining the reliability of these citation sources largely remains a manual task for subject experts \cite{redi2019citation}. Prior work has shown that novices rely on superficial criteria for judging the trustworthiness of web-based information and sources, while suggesting that domain knowledge plays an important role in how people evaluate provenance information where greater domain knowledge leads to a more reliable selection of sources~\cite{brand2017source}. DoubleCheck supports people with different expertises and backgrounds by integrating domain knowledge of historical scholarship (i.e., of the American Civil War era) to organize sources according to their trustworthiness, while providing actionable visualizations to indicate the credibility of an ID.  

\section{Civil War Photo Sleuth: Background, Challenges, and Design Goals} \label{limitationscwps}

The American Civil War (1861-65) is considered to be one of the earliest major conflicts to be widely photographed. More than 3 million soldiers fought in the war and most of them were photographed at least once. While museums, libraries, and personal collections have helped preserve a lot of these photographs over the years, only 10--20\% of them are identified. Civil War photography has generated a lot of interest among historians, collectors, dealers, genealogists, archivists, and other experts, who often try to identify unknown photos for personal, cultural, and economic reasons. However, the identification process is complex and challenging, and the methods they employ are largely manual -- which often involves identifying visual clues in photos and manually analyzing hundreds of low-resolution photographs, military records, and reference books to corroborate evidence. 

To address this problem, Mohanty et al.~\cite{mohanty2019photo} developed Civil War Photo Sleuth (CWPS), a free, public website where users can identify unknown Civil War portraits using a novel person identification pipeline that combines the strengths of crowdsourced human expertise and AI-based facial recognition. The pipeline draws analogies to finding a needle in a haystack, and has three main components: 1) upload and tag the unknown photo with visual clues such as uniform details (i.e., coat color, chevrons, shoulder straps, etc.), 2) use the uniform clues to generate search filters based on service records (i.e., a dark coat color means a Union soldier), and 3) use facial recognition to retrieve facially similar-looking search results from a pool of potential candidates that satisfy the search filters. 

This ‘haystack model’ is designed to prevent misidentifications by not allowing the algorithm to automatically pick the top-ranked result as the best match, nor does it display the algorithm’s quantitative similarity confidence scores. Instead, it adopts a human-led, AI-supported approach where the user carefully analyzes the search results for a potential match based on not just facial similarity, but also whether biographical details (i.e., service records, location, etc.) align or not. Once the user identifies a match, CWPS links the face and identity together and displays the ID on a dedicated page for the photo. 

Initial evaluation of the platform after one month of being released to the public showed that CWPS was successful in helping users identify unknown photos. 119 newly identified photos were matched to 88 identities in the first month, of which 13 were found to be incorrect. Mohanty et al.~\cite{mohanty2019photo} categorized these incorrect IDs as the hardest type of identification -- photos that did not have any prior name inscriptions, nor an exact matching view, called a replica, in the database. A follow-up benchmarking of the underlying facial recognition \cite{mohanty2020photo} showed that it was prone to false positives (i.e., retrieved a large number of search results for the user to filter out and find the correct match for the unknown photo), leading to the possibility of automation bias playing a role in misidentifications happening on the platform. The platform’s open participation model, without any validation measures, has raised concerns about the trustworthiness of IDs proposed on CWPS and an increase in misidentifications as the site grows in size \cite{harris2019civil}. 

CWPS originally allowed users to identify photos in a single-step, comparison interface where they can indicate a match by clicking an “Identify” button.  In order to foster more accurate, deliberate decision-making, CWPS developers subsequently released a feature that allowed users to make fine-grained photo identification decisions in two-steps: photo comparisons followed by photo identification. Users first compare two photos and indicate whether the query photo and target photo(s) are replicas of each other, or show a facial match (same person, different views), or show different people. Once the photo comparison is complete, the users can then factor in additional biographical information (along with the facial similarity information from the previous step) to indicate how confident they are about the query photo being the target ID. These photo comparison and identification confidence scores are then displayed on the photo page in the form of bar chart visualizations. 

While this validation workflow introduced a form of community stewardship on CWPS, there still remained gaps in terms of assessing the quality of photo identifications made on CWPS. In the following sections, we discuss specific challenges that can potentially lead to inaccurate assessment of the photo IDs, and how we address them through the design of DoubleCheck. 

\subsection{Challenge: Inadequate Provenance Information} 

Provenance is an important aspect of history research, especially for generating evidence to support historical claims \cite{levine2008importance,schellenberg1965principle}. Here, in the context of historical photo identification, rich provenance information can instill confidence in users about the identification of a photo. Prior work has also shown provenance information to be useful for improving users’ assessment of user-generated data in online communities~\cite{forte2014designing}. 

CWPS currently collects and displays two types of provenance information: 1) \textit{photo source}, i.e., where the user found the photo; and 2) \textit{biography source}, i.e., where the user found the service records or biographical information of the person in the photo. However, it does not explicitly collect any information about how or where the user discovered the identity of the person in the photo (e.g., ``according to book ABC, the person in this photo is John Doe''). 

CWPS also does not make any distinction between such identifications where the user knew the identity prior to uploading the photo (i.e., pre-identified photos \cite{mohanty2019photo}) versus new identifications where the user discovered the photo's identity by using the website (i.e., post-identified photos \cite{mohanty2019photo}). If users already know the identity of a person from an external source prior to uploading the photo, they do not have the option to provide that information to the system prior to viewing the search results. If they find a facial match among the search results for the right person, they can click the "Identify" button and the identity will get linked to the photo. However, this external source information about the photo’s identity is not captured at all. This is a critical piece of information for users trying to assess an identification, and researchers evaluating the system’s effectiveness.

Consider the following examples: 

\begin{figure}[htbp!]
    \centering
    \includegraphics[scale=0.20]{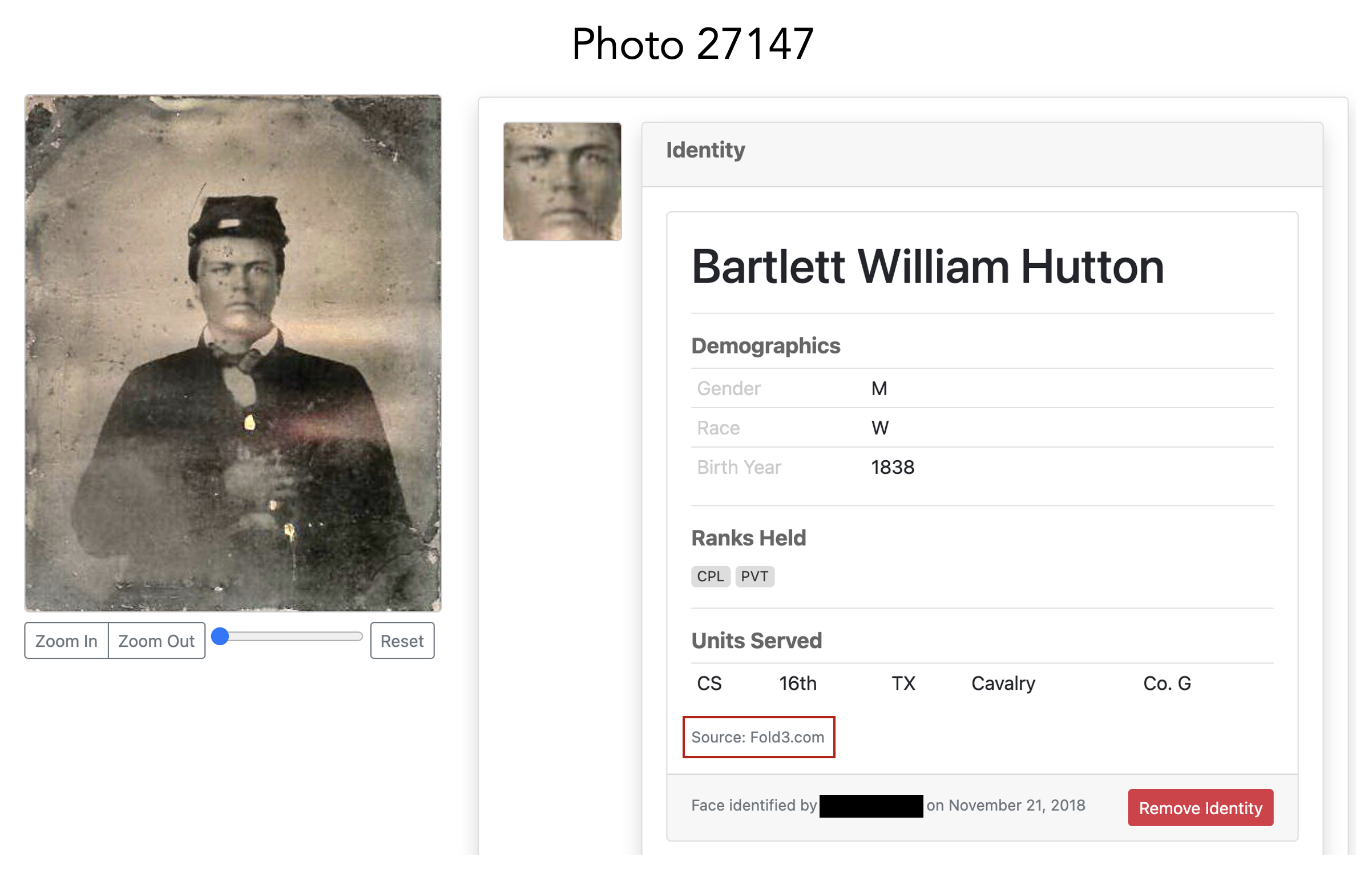}
    \caption{Photo Page on the original CWPS Interface. Photo 2714 identified as Bartlett William Hutton, with Fold3 as the biography source.}
    \label{fig:fold3}
\end{figure}

\paragraph{\textbf{Example 1:} A user identified Photo 27147 as Bartlett William Hutton (see Figure~\ref{fig:fold3}). He served as a private and a corporal in Company G of the 16th Texas Cavalry regiment. These service records were added by the user, who cited Fold3.com as the source of this information (biography source). If someone wants to cross-check the accuracy of this information, they may check Fold3 for that particular soldier’s information. However, Fold3 may only be useful for tracing back the biographical details, and may not say anything about Photo 27147 being Bartlett William Hutton.\\\\}

\begin{figure}[htbp!]
    \centering
    \includegraphics[scale=0.20]{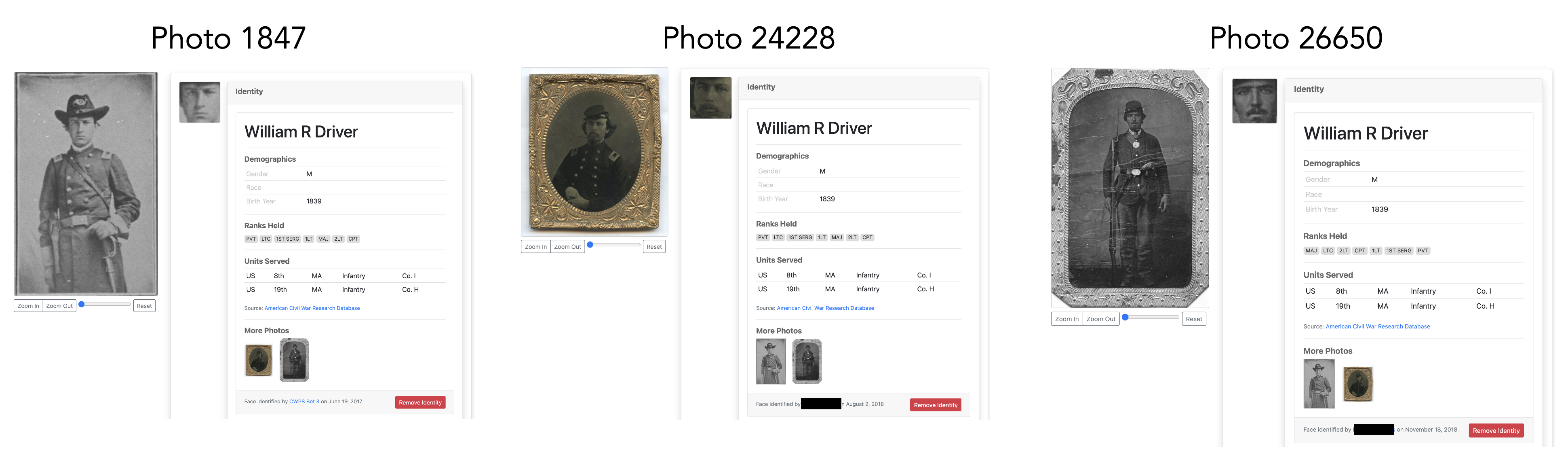}
    \caption{Photo Page on the original CWPS Interface. Photo 1847, Photo 24228 and Photo 26650 have all been identified as William R Driver. It is unclear from the photo page whether users knew the IDs of Photo 24228 and Photo 26650.}
    \label{fig:williamdriver}
\end{figure}

\paragraph{\textbf{Example 2:} Photo 1847 (see Figure~\ref{fig:williamdriver}), one of the original seeded photos on the website (from the MOLLUS collection), is identified as William R Driver, who served in the 8th and 19th Massachusetts Infantry regiment. One user identified Photo 24228 (see Figure~\ref{fig:williamdriver})) as the same person. Another user identified Photo 26650 (see Figure~\ref{fig:williamdriver})) also as the same person. However, from the information on the photo page, it is not clear whether those users knew the identities of Photo 24228 and Photo 26650 from external sources prior to uploading, or if they identified the photos solely on the basis of facial similarity.\\\\}

While the biography source may sometimes include provenance information about the photo’s identity, oftentimes it only includes text-based biography and service records with no specific reference to a photo. Similarly, the information on the photo page does not indicate whether a photo was identified solely on the basis of facial similarity, or if it is supported by external source information. Without a comprehensive provenance, users may find it difficult to accurately assess whether an identification is airtight or not. For the same reason, history scholars may not be able to use photos identified on CWPS for their research purposes if they are unable to verify the accuracy of an identification.  

\subsubsection{Design Goal 1 (Information Provenance):} In order to support accurate assessment of photo IDs, users should be able to access comprehensive provenance information for the IDs, while being able to distinguish between pre-identified and post-identified photos on CWPS. This design goal aligns with the provenance aspect of the \emph{assessable design} framework~\cite{forte2014designing} which suggests being transparent about where the information came from. 

\subsection{Challenge: Unknown Credibility for Photo Identifications}

\begin{figure}[htbp!]
    \centering
    \includegraphics[scale=0.20]{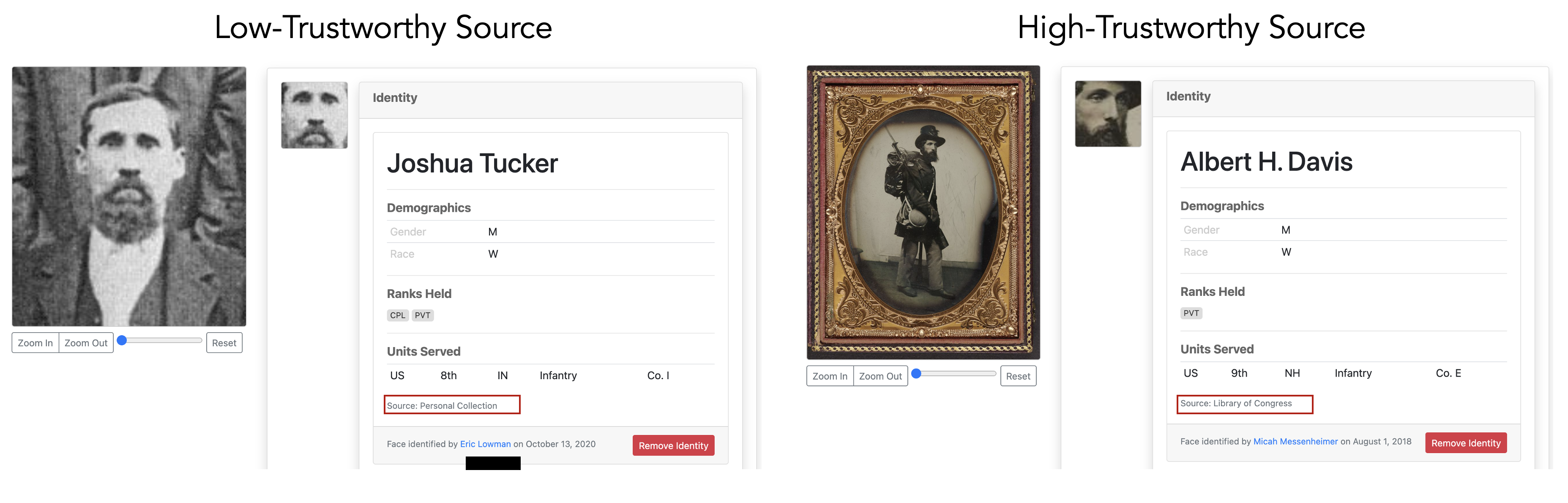}
    \caption{Photo Page on the original CWPS Interface. Two different IDs with different source trustworthiness. The interface does not make any distinction between the two.}
    \label{fig:lowhightrustworthy}
\end{figure}

The accuracy of photos identified on CWPS may vary due to multiple factors. One factor is the reliability of provenance information. Some sources may be considered more reliable and trustworthy than others. For example, a primary source, such as a period inscription on the photo, is more reliable than a word-of-mouth ID coming from a dealer or descendant \cite{luther_trustworthiness}. Photos may also vary in terms of information completeness, which affects how users assess the quality of the identification. An identification missing critical information pieces, such as the source or service records, may be more difficult to verify than an identification with complete information. However, the CWPS system does not factor in this information for assessing, or indicating, the quality of an identification on the photo page. As a result, there is no difference in the photo page of an identification with a low trustworthy (or missing) source versus an airtight identification (see Figure~\ref{fig:lowhightrustworthy}). 

A photo identification may also require additional verification if it lacks credible evidence. However, the quality of an identification is subject to change if new corroborating information (e.g., new sources, additional photos, etc.) is presented \cite{luther2016fellow}. \vm{Further, because users on CWPS conduct original historical research, they may propose more than one identity for a given photo based on the available evidence~\cite{10.2307/27172788}.} Since CWPS does not display any quality indicators, there is no such distinction between the different stages of identity verification\vm{, nor any way to resolve cases of conflicting identities even if the evidence favors one over others.}

Historians, with their years of training and expertise, may have a better understanding of domain knowledge and different identification sources, and therefore, may assess photo identifications differently from novices. Without any indicators for identification quality or source trustworthiness, there is a possibility that novice users may also incorrectly assess the accuracy of an identification. 

While the community's votes on a photo ID can qualify as an acceptable signal for an ID's credibility, relying solely on that signal can make one prone to \vm{"groupthink"~\cite{muchnik2013social}}. This necessitates designing a robust quality assessment mechanism, which can reflect both the community's opinions and available provenance information, while providing reliable indicators to help assess the quality of a photo ID. 

\subsubsection{Design Goal 2 (Information Stewardship):} Following the \emph{assessable design} framework recommendation for information stewardship~\cite{forte2014designing} --- \emph{help people understand how the information is maintained} --- the platform should factor in the community's opinions and trustworthiness of involved sources to provide easy-to-understand, actionable indicators to help users  assess whether a given photo ID can be verified or not.

\section{System Description: DoubleCheck}

We developed \emph{DoubleCheck} for helping users assess identities accurately on CWPS. To address the challenges discussed in Section \ref{limitationscwps}, we built an architecture to support provenance and stewardship as part of DoubleCheck, which required us to modify the underlying data architecture of how photos and identities on CWPS are stored and linked to each other. 

We re-designed the photo page of CWPS with a vertical layout; the top section displays the photo, the identified name and the metadata information (photo source, owner, inscription and photographer), the next section displays the face and all the identities proposed (see Figure \ref{fig:allids}) and finally, the bottom section displays an activity feed of all user activities. In this section, we will be focusing on the identities section, which displays the provenance and stewardship visualizations.

\subsection{Provenance}

CWPS already captures and displays two types of provenance information: \emph{photo sources} and \emph{biography sources}. In order to help users accurately assess a photo’s identity, we built a workflow to capture a third, new form of provenance, \emph{identification sources}, i.e., evidence that suggests who the person in the photo is (\emph{Design Goal 1}). 


\subsubsection{Source Categories and Trustworthiness}

To integrate information about source trustworthiness, we needed a predefined pool of source types. We scraped and analyzed all the biographical sources that users had previously provided while identifying photos on CWPS. We iteratively coded 1326 unique biographical sources into high-level \emph{source types}, creating a pool of common sources. For example, "Yale Beinecke Library" and "Maine State Archives" were categorized as a "website run by a library, archive, or a museum." Sources that were frequently cited by users became their own source types, such as Library of Congress, Ancestry.com, etc. Table~\ref{tab:source-types} shows the final list of source types that will now be used for citing identification sources on CWPS.


\begin{table}[htbp!]
\tiny
\begin{tabular}{|l|c|
>{\columncolor[HTML]{EFEFEF}}c |}
\hline
Period Inscription with Valediction &
   &
  \cellcolor[HTML]{EFEFEF} \\ \cline{1-1}
Period Inscription without Valediction &
   &
  \cellcolor[HTML]{EFEFEF} \\ \cline{1-1}
Period Inscription on Union Case &
   &
  \cellcolor[HTML]{EFEFEF} \\ \cline{1-1}
Period Inscription on Album Page &
  \multirow{-4}{*}{\textit{Inscription}} &
  \multirow{-4}{*}{\cellcolor[HTML]{EFEFEF}\textbf{Primary Source}} \\ \hline
\begin{tabular}[c]{@{}l@{}}Period publication\\  (e.g. regimental history, newspaper)\end{tabular} &
   &
  \cellcolor[HTML]{EFEFEF} \\ \cline{1-1}
\begin{tabular}[c]{@{}l@{}}Modern publication \\ (e.g. book, article)\end{tabular} &
  \multirow{-2}{*}{\textit{Books}} &
  \cellcolor[HTML]{EFEFEF} \\ \cline{1-2}
\begin{tabular}[c]{@{}l@{}}Period documents (e.g., letter, album, diary, \\ obituary, burial card, pension records, discharge papers)\end{tabular} &
  \textit{Groupings} &
  \cellcolor[HTML]{EFEFEF} \\ \cline{1-2}
Library of Congress &
   &
  \cellcolor[HTML]{EFEFEF} \\ \cline{1-1}
National Archives &
   &
  \cellcolor[HTML]{EFEFEF} \\ \cline{1-1}
\begin{tabular}[c]{@{}l@{}}US Army Heritage and Education Center\\ (MOLLUS)\end{tabular} &
   &
  \cellcolor[HTML]{EFEFEF} \\ \cline{1-1}
Other library, museum or archive &
  \multirow{-4}{*}{\textit{\begin{tabular}[c]{@{}c@{}}Scholarly \\ Website\end{tabular}}} &
  \multirow{-7}{*}{\cellcolor[HTML]{EFEFEF}\textbf{\begin{tabular}[c]{@{}c@{}}Secondary Sources \\ (Scholarly)\end{tabular}}} \\ \hline
Modern Inscriptions &
  Inscription &
  \cellcolor[HTML]{EFEFEF} \\ \cline{1-2}
Ancestry.com &
   &
  \cellcolor[HTML]{EFEFEF} \\ \cline{1-1}
Fold3 &
   &
  \cellcolor[HTML]{EFEFEF} \\ \cline{1-1}
Find A Grave &
   &
  \cellcolor[HTML]{EFEFEF} \\ \cline{1-1}
American Civil War Research Database (HDS) &
   &
  \cellcolor[HTML]{EFEFEF} \\ \cline{1-1}
Other genealogy website &
  \multirow{-5}{*}{\textit{\begin{tabular}[c]{@{}c@{}}Genealogy \\ Website\end{tabular}}} &
  \cellcolor[HTML]{EFEFEF} \\ \cline{1-2}
Auction house website &
   &
  \cellcolor[HTML]{EFEFEF} \\ \cline{1-1}
eBay listing &
  \multirow{-2}{*}{\textit{Auction}} &
  \cellcolor[HTML]{EFEFEF} \\ \cline{1-2}
Dealer or collector &
   &
  \cellcolor[HTML]{EFEFEF} \\ \cline{1-1}
Family or descendant &
  \multirow{-2}{*}{\textit{Word-of-Mouth}} &
  \cellcolor[HTML]{EFEFEF} \\ \cline{1-2}
Misc. Websites / social media (e.g. Facebook, Blogs) &
   &
  \cellcolor[HTML]{EFEFEF} \\ \cline{1-2}
Other &
   &
  \multirow{-12}{*}{\cellcolor[HTML]{EFEFEF}\textbf{\begin{tabular}[c]{@{}c@{}}Secondary Sources \\ (Non-Scholarly)\end{tabular}}} \\ \hline
\end{tabular}
\caption{List of different source types with their categories. Primary sources are considered the most trustworthy, followed by scholarly secondary sources. Non-scholarly secondary sources are the least trustworthy.}
\label{tab:source-types}
\end{table}

According to a survey of identification sources for Civil War portraits \cite{luther_trustworthiness}, the most trustworthy sources were primary sources (i.e., sources that were present physically on or near the photo), followed by scholarly secondary sources (i.e., sources that relied on the expertise of published authors and museum professionals). Least trustworthy were other (non-scholarly) secondary sources, e.g., word-of-mouth identifications made by dealers or descendants or from social media. On CWPS, if users do not provide any source types, we consider them in the same bracket as a word-of-mouth identification (a low trustworthy source). Based on this survey, we categorized the different source types as either primary, scholarly secondary or non-scholarly secondary sources (see Table~\ref{tab:source-types}). The system factors in this information about source trustworthiness for assessing the quality of photo identifications, which we discuss in Section~\ref{qualityassessment}.

\subsubsection{Linking Sources and Identities}\label{linkingsources}

We modified the underlying data architecture of CWPS to link identification sources with photo ID. Each source is denoted by a source type (see Table~\ref{tab:source-types}) along with details of the source. For example, if a user wants to add a recently published book as an identification source, they can add "Modern Publication" as the source type, and the name of the book as the source details.

\begin{figure*}[t!]
    \centering
    \begin{subfigure}[t]{0.5\textwidth}
        \centering
        \includegraphics[height=1.5in]{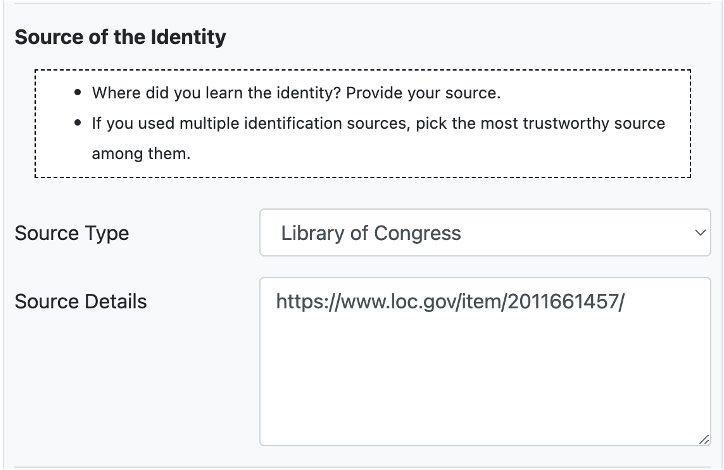}
        \caption{UI for providing ID source information (source type and details)}
    \end{subfigure}%
    ~ 
    \begin{subfigure}[t]{0.5\textwidth}
        \centering
        \includegraphics[height=3in]{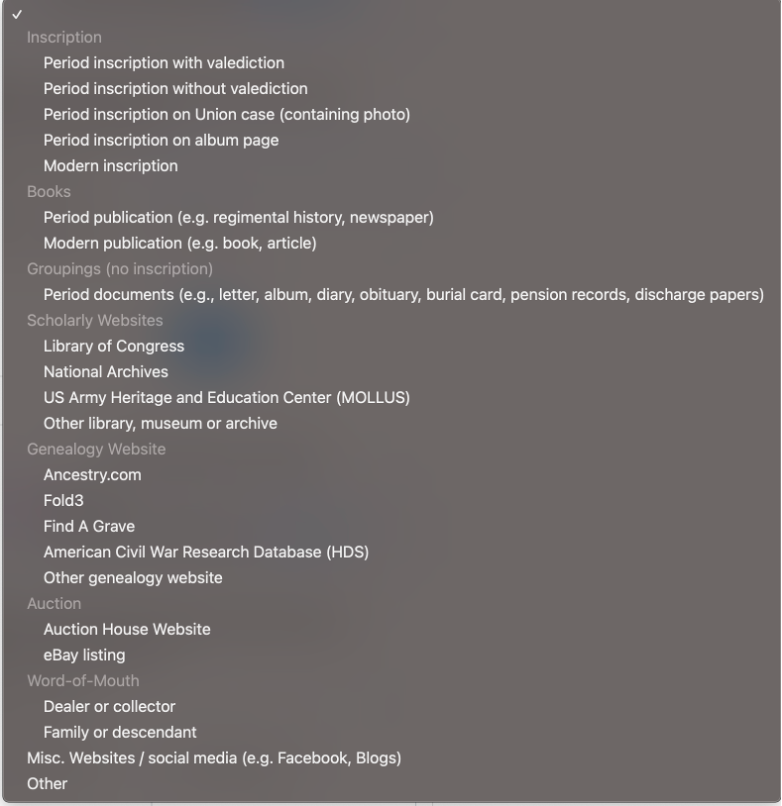}
        \caption{Dropdown menu showing a list of source types}
    \end{subfigure}
    \caption{Capturing Provenance Information}
    \label{fig:provenancecapture}
\end{figure*}

For users who know the identity of the photo prior to uploading the photo, we added an additional pre-identification step at the beginning of CWPS's identification workflow. This step captures the user's prior knowledge about the photo's identity (Design Goal 1), including the source information (see Figure~\ref{fig:provenancecapture}). Once the face is identified, this original identification source gets linked to the photo ID. 

If a user does not know the identity of a photo beforehand, they can skip the pre-identification step and go directly to the search results page for checking the facially similar search results. If they find a match, the system infers that the query photo was identified on the basis of facial similarity with the target photo and the biographical details lining up. Here, the target face and target identity are both linked to the query face, and the target photo becomes a source for identifying the query photo as the target identity. In other words, the original identification source used to identify the target photo becomes an additional (indirect) source for the query photo. 

This workflow ensures that all identification sources are captured for both pre-identified and post-identified photos. Since a single face can be linked to multiple faces for a specific identity, this workflow also supports multiple identification sources for a photo identification.

\subsubsection{Visualizing Provenance}

As part of the DoubleCheck framework, each identity now has an identification sources section (\emph{Design Goal 1}), where all identification sources are listed under different subsections representing the source categories: Primary, Secondary (Scholarly), and Secondary (Others).

\begin{figure}
    \centering
    \includegraphics[scale=0.27]{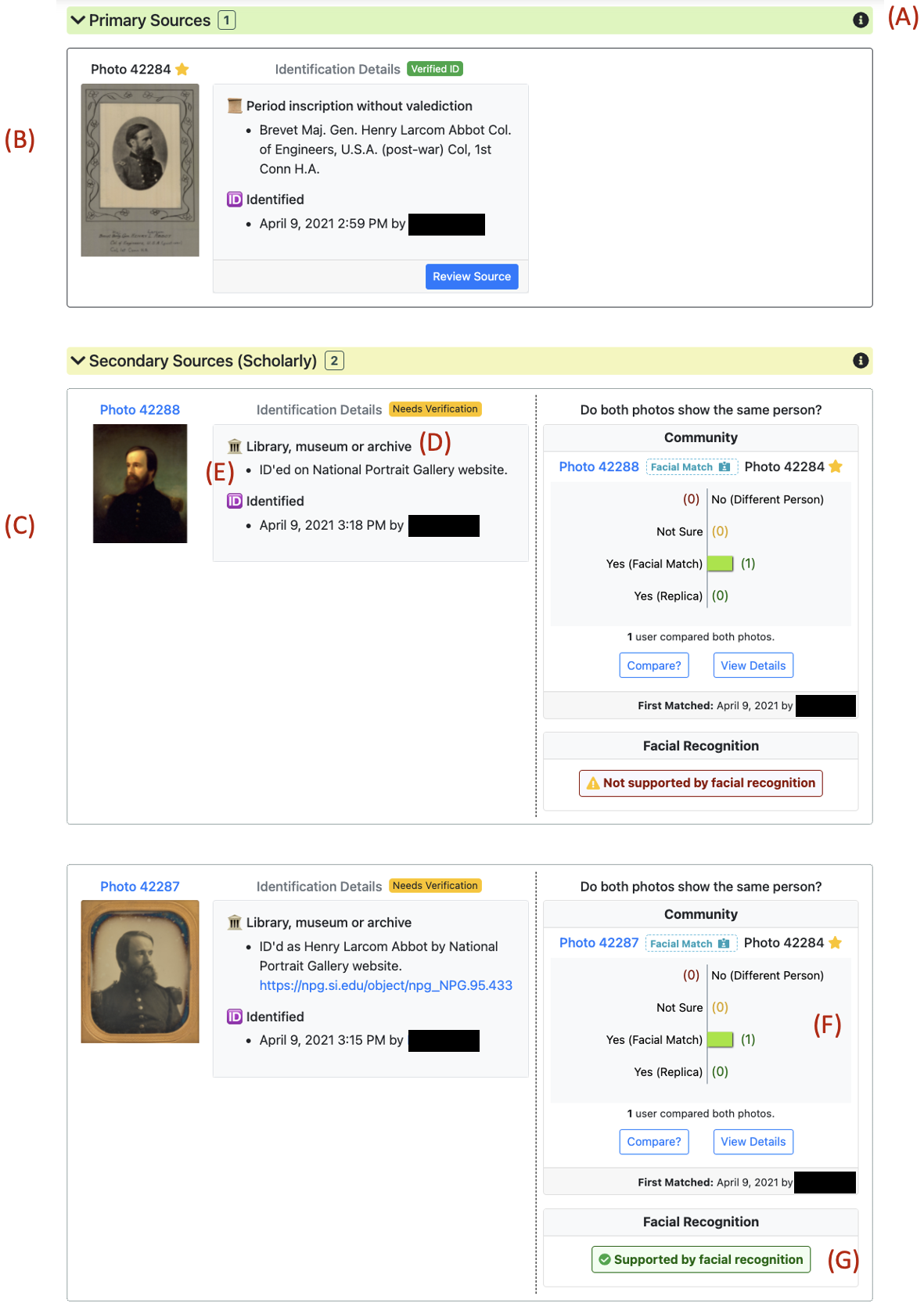}
    \caption{DoubleCheck's presentation of provenance information on the photo page. 
    \textbf{(A)} All identification sources are organized into high-level source categories, in the order of their trustworthiness: Primary, Secondary (Scholarly) and Secondary (Non-Scholarly Sources).
    \textbf{(B)} A source that came with the query photo. 
    \textbf{(C)} A source that was linked to the query photo through facial similarity. 
    \textbf{(D)} The source type (see Table~\ref{tab:source-types}). 
    \textbf{(E)} The source details. 
    \textbf{(F)} Community's opinion on the visual match type (i.e., whether the source photo and the query photo show the same person or not). 
    \textbf{(G)} Badge displaying whether facial recognition considers the source photo and the query photo to be the same person or not. 
    }
    \label{fig:allidsources}
\end{figure}

Each identification source is represented by the following elements (see Figure~\ref{fig:allidsources}): 
    \begin{itemize}
        \item[1.] the source photo, which can either be the query photo or another photo of the person that is linked to the query photo, 
        \item[2.] the source type (see Table~\ref{tab:source-types}) and details that are linked to the source photo,
        \item[3.] the user who identified the source photo,
        \item[4.] source matching and reliability information, and
        \item[5.] the user who matched the source photo with the query photo, in case of a source that was linked to the query photo via facial similarity.
    \end{itemize}

The source categories are ordered by trustworthiness; primary sources first, followed by scholarly secondary sources, and finally other secondary sources. The subsection headings indicate the category type, and tooltips provide additional information. 
    
If a photo is pre-identified, then one of the sources is represented by the photo itself. If a photo has been linked to another photo for the same identity, they both appear as identification sources on each other’s photo page. 

\subsection{Stewardship}

CWPS's photo ID validation workflow (see Figures~\ref{fig:comparisonstep1} and~\ref{fig:comparisonstep2} in Appendix) -- \emph{users comparing photos for facial similarity and voting on photo IDs} --  already encapsulated a form of \emph{community stewardship}. 

The community's opinions for the facial comparison between two matched photos, also depicted in the form of a bar chart (see Figure~\ref{fig:allidsources} (f)), conveys the reliability of the facial similarity. In addition to that, a badge indicating whether facial recognition supports the match or not is also displayed. However, with the introduction of DoubleCheck's provenance feature, the matched photo's original ID source also becomes an additional (indirect) source for the query photo, and therefore, the facial similarity stewardship now gets a new meaning --  it essentially provides information about whether a linked photo can be reliably used as a source for corroborating the query photo's ID. 

For example, consider a query photo and a photo linked to it, where the linked photo has a period inscription (a highly trustworthy primary source). If the majority of users consider both photos to be facial matches or replicas, and it is supported by facial recognition, one might have high confidence on the proposed ID due to the presence of a primary source linked to it. However, if the community increasingly considers both photos to show different people, the proposed ID cannot be supported by the linked primary source with a high degree of confidence.  

The community's confidence votes on a given photo ID, displayed in the form of a bar chart (see Figure~\ref{fig:idreliabilityvisalizations} in Appendix), serve as an indicator for the ID's reliability. However, this stewardship visualization does not directly convey anything about the ID's reliability nor does it factor in provenance information; users have to interpret the available data and make their own assessment about the ID. DoubleCheck builds upon this existing stewardship, and combines it with the source trustworthiness information to propose a new form of information stewardship: a \emph{quality assessment engine} to indicate whether an ID is verified or not.

\begin{figure}
    \centering
    \includegraphics[scale=0.30]{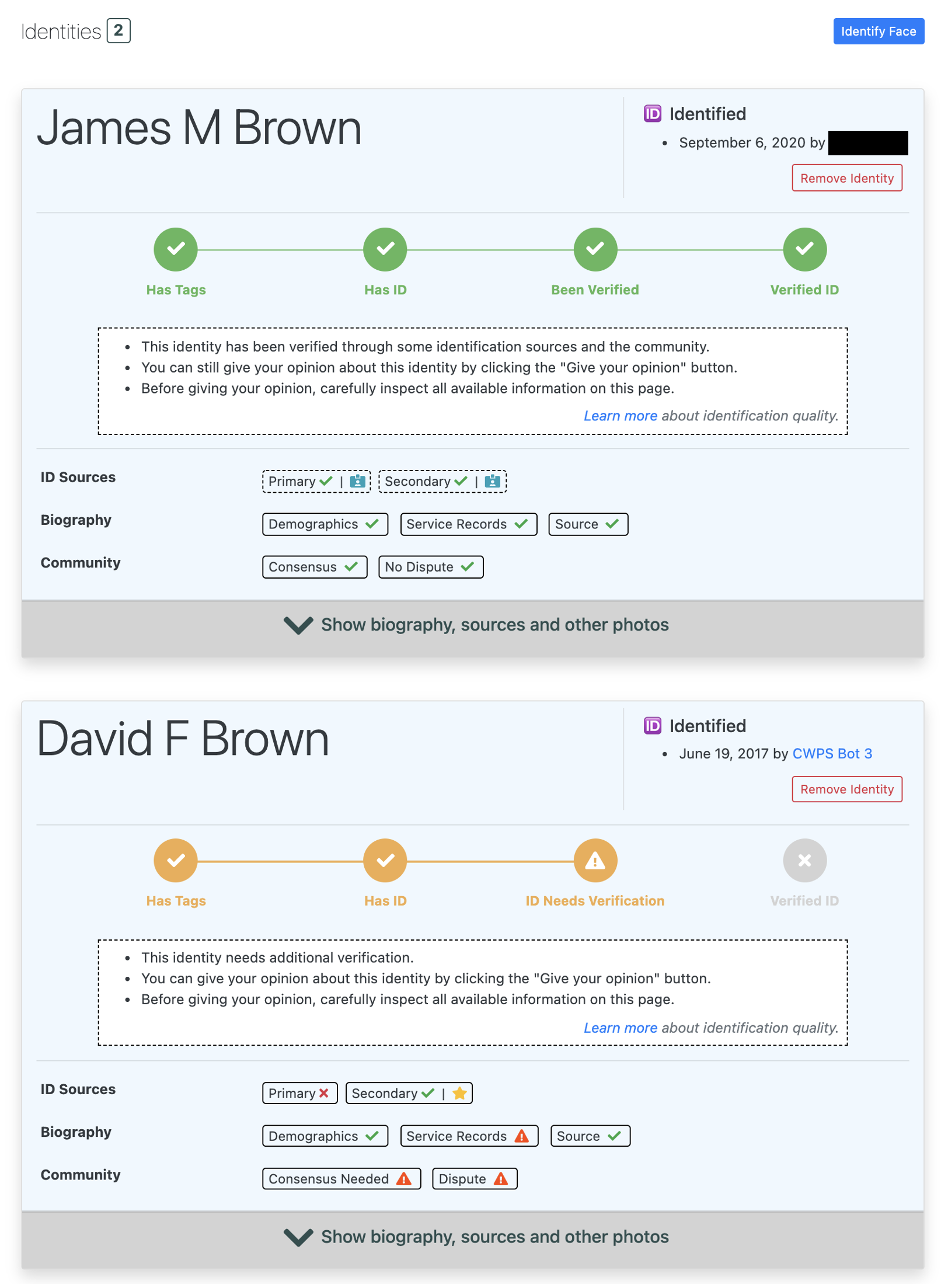}
    \caption{Overview of all identities proposed for a photo. Here, the photo has an ID conflict (i.e., multiple IDs have been proposed). Each identity card expands to show the biography, community's opinions, identification sources and other photos. The quality assessment visualization shows the current status of the ID. Here, James M Brown is a \textit{verified ID} whereas David F Brown \textit{needs verification}. The overview badges show whether there is consensus and/or dispute among the community. James M Brown has received enough votes from the community without any dispute, whereas the community disagrees with David F Brown.}
    \label{fig:allids}
\end{figure}

\subsubsection{Quality Assessment Engine} \label{qualityassessment}

We built an automated quality assessment engine and visualization to draw the community’s attention to the identification quality, and encourage them to find additional evidence and perform necessary tasks to improve the quality of identifications (\emph{Design Goal 2}). This engine is based on provenance and community stewardship data, and factors in both source trustworthiness and the community’s expertise. 

Every photo on CWPS goes through a sequential four-step quality assessment process, starting with (1):

\begin{enumerate}
    \item \textbf{Needs Tags:} A newly uploaded photo is assigned this badge automatically, representing the lowest quality level. When the user adds the minimum required tags for photo metadata and uniform evidence information, the quality badge is upgraded automatically to "Needs ID".

    \item \textbf{Needs ID:} Photos that satisfy the "Needs Tags" criterion and are unidentified are assigned this badge. When at least one identification is proposed for photos with this badge, the badge is upgraded to "Needs Verification."

    \item \textbf{Needs Verification:} Identified photos that satisfy the "Needs Tags" criterion are assigned this badge. They require additional verification, either through community consensus, new identification sources, or both. Specifically, the assessment system checks for the following conditions:
    
    \begin{itemize}
        \item the photo came with a primary identification source and there is no community dispute,
        \item the photo came with a secondary scholarly source and there is community consensus,
        \item the photo was identified as a replica of another photo whose identity has already been verified, or 
        \item the photo was identified as a facial match of another photo whose identity has already been verified, and there is community consensus and no dispute.
    \end{itemize}
    
    If any one of these conditions is satisfied, the quality badge is upgraded to "Verified ID". If none of these conditions is met, then the photo remains in the "Needs Verification" category awaiting further validation. 

    \item \textbf{Verified ID:} If an identity clears the assessment check conditions in the "Needs Verification" stage, then it becomes a "Verified ID," the highest quality level.

\end{enumerate}

Depending on which the step the photo is currently in, it is automatically assigned a quality assessment badge. The assessment engine automatically checks for these conditions every time there is a new vote or when a source is added or removed and updates the badges accordingly. In case of conflicting identities for the same photo, we use these checks to pick a "winning" identity. 

We display these badges next to the identity title for a photo. We also display a visualization that shows the progression of a quality for an identification (see Figure~\ref{fig:allids}). Each of these badges is instructive in nature, and the visualization displays instructions for how to proceed to the next step based on the system’s current assessment of the identification. For example, an identity that "needs verification" requires users to pay attention to the information on the page and also to give their opinion, which can help in verifying or debunking the identity. 

Users can not only use these badges to form an assessment of a given photo, but also extrapolate their assessment to other photos (with the same identity) that have been matched to it. For instance, if they are confident about their assessment of a particular photo’s identity, they will be able to extend the same assessment for a replica unless there is some other information conflict. In such cases, the assessment engine automatically assigns the same badge to the replica as the original photo unless there is a dispute. However, users may more carefully assess a facial match since there is a stronger possibility of incorrect matches.

\section{Evaluation}

We obtained permission to publicly launch DoubleCheck on CWPS in the last quarter of 2020. 
We conducted an exploratory evaluation study to understand how well users with different expertise levels could assess and validate Civil War photo IDs using CWPS with DoubleCheck. Specifically, we wanted to understand 1) how the provenance and stewardship visualizations (i.e., quality assessment badges) impacted users’ assessment of an ID. Even though the evaluation is a mixed-methods study, it is largely a qualitative user study, aided by some quantitative analysis of the logs. The entire study was approved by our university's IRB.

\subsection{Log Analysis}

To understand how the community contributed towards the provenance and stewardship visualizations, we examined website logs for all user activities for twelve months since the launch of the new features. For each identified photo, we also analyzed the identification sources provided for it, how many sources were provided, and whether the source was a pre-identified source or linked via facial similarity. We further checked the completeness of the source information (i.e., whether the source type and details were provided).

We also analyzed the quality assessment badges of all the photos on CWPS. When we launched the new system in 2020, none of the photos had a "Verified ID" badge, as identities with period inscription sources were not automatically categorized initially. For photos and identifications that underwent a change in the quality badge, we analyzed the community actions that resulted in this change. We only focused on the photos that were identified during this twelve month period.

\subsection{Lab Study}

In order to understand how well the DoubleCheck version of CWPS supports diverse users in assessing and validating the quality of photo identifications, we also conducted an exploratory lab study. 

\subsubsection{Participants}

We recruited 15 participants representing the three major expertise levels and our goals for the new features: 

\begin{description}

    \item[Students:] Undergraduate and graduate (master’s) students concentrating in history who use Civil War photos for their coursework and research projects, but are not (yet) employed in a professional capacity as historians. We recruited five students via recommendations from our university's history department. None of the students had used CWPS before, or were known to the authors prior to the study. Three students were men and two were women, and all were in the "18 to 30" age group. We anonymize them with identifiers S1--S5.
    
    \item[Amateur Experts:] Experienced users of Civil War Photo Sleuth who have added over 50 photos each and have substantial knowledge of Civil War history, but are not professional historians. We recruited five amateur experts from the CWPS contact list. All five users were men, and they were distributed across different age groups (two in "18 to 30", two in "31 to 40", and one in "51 to 60"). We anonymize them with identifiers C1--C5. C1 and C3 are among the most active daily users on CWPS. Only two of the five had used DoubleCheck before. 
    
    \item[Historians:] Expert historians with a graduate degree in history, specializing in American Civil War history, but with little or no previous experience with CWPS. We recruited five historians via recommendations from our university's history department. Three historians were men and two were women. They were distributed across different age groups (two in "18 to 30", two in "31 to 40", and one in "51 to 60"). We anonymize them with identifiers H1--H5. None of them had used DoubleCheck before. 
    
\end{description}

\subsubsection{Dataset}

For the study, we created a dataset of 10 different photos identified on CWPS. Three of these photos had an ID conflict, i.e., multiple identities were proposed. For two of these photos, one ID was correct and the other one was incorrect; the correct photo had a verified ID. The community had already researched both photos, voted on the correct ID, and left credible evidence in the comments. Both IDs had verifiable sources and were linked to additional photos as well. The third photo was one of the seeded photos on CWPS, but was originally misidentified. We added another false ID, making both IDs for the third photo incorrect. 

The remaining seven photos only had one ID. Four of these photos had a "Verified ID" badge, and three of them had a "Needs Verification" badge. During the course of the study, one of these IDs received enough votes and got verified. All photos had multiple sources; eight of them had photo sources linked via facial matches, while two of them had replica photo sources. Some of these photos had period inscriptions.

\subsubsection{Procedure} 

The entire study was conducted online via recorded Zoom sessions, with at least one researcher attending each session. Each participant first completed a consent form and a pre-survey describing their demographics and Civil War photography experience. Participants were randomly assigned three photos from the dataset. In order to understand the impact of the quality assessment badges, we ensured that they received one photo with a "Verified ID" and one photo with an ID conflict. 

In order to understand whether the new \textit{DoubleCheck} version of the website improved over the original CWPS system, we used a within-subjects, comparative study design. Each participant first reviewed these photographs one-by-one in the original CWPS system using a think-aloud protocol. After each photo assessment, they completed an online survey about the extent to which they found that the information they saw was complete, verifiable, trustworthy, and accurate on a four-point Likert scale (strongly disagree, slightly disagree, slightly agree, and strongly agree), similar to the metrics used in \cite{forte2014designing}. We then asked some semi-structured interview questions about the experience.

Next, the participant checked the same photos in the new \textit{assessable} interface, one at a time, using a think-aloud protocol. After each photo assessment, participants completed a survey about the information they encountered, and answered how it affected their original assessments about the information (i.e., completeness, verifiability, trustworthiness, and accuracy). They answered using a five-point Likert scale (strongly lowered, slightly lowered, did not affect, slightly raised, and strongly raised) (see Appendix). Finally, the participant completed a summative post-survey of standard usability questions (e.g., ease of use, usefulness of features, instruction clarity, preferred system, etc.)  and answered some semi-structured interview questions about their overall opinions of both the interfaces. 

We maintained this sequence (original CWPS first, CWPS with DoubleCheck second) for all the participants, rather than using a randomized sequence, as we did not want participants' assessments to be biased in favor of DoubleCheck after seeing additional information in the new interface. This design allowed us to observe if the original interface misled the participants towards incorrect assessments, and if, subsequently, the assessable interface helped correct them. The participants were asked to assess the ID using only the information available to them in the interface (i.e., they could click the URLs, if available). 

\subsubsection{Data Analysis}
The first author fully transcribed and analyzed the interviews and think-aloud recordings using an inductive qualitative thematic approach \cite{braun2006using} . The transcript sections were first divided according to the interface in question (i.e., original CWPS or DoubleCheck), followed by an open coding of the transcripts using MAXQDA 2020 \cite{verbi_software_maxqda_2019}. The first author iterated and settled on a total of 28 codes (e.g., change in opinions, comparison interface, source trustworthiness, etc.) for 634 coded segments across all the transcripts. These codes were then organized into themes as described in Section~\ref{sec:findings} after discussing with the co-author. 



\section{Findings}\label{sec:findings}

Using the methods above, we evaluated how well DoubleCheck supported CWPS users in assessing photo identifications, compared to the original version of CWPS. 

\subsection{Assessability Outcomes}

\begin{figure}[htbp!]
    \centering
    \includegraphics[scale=0.21]{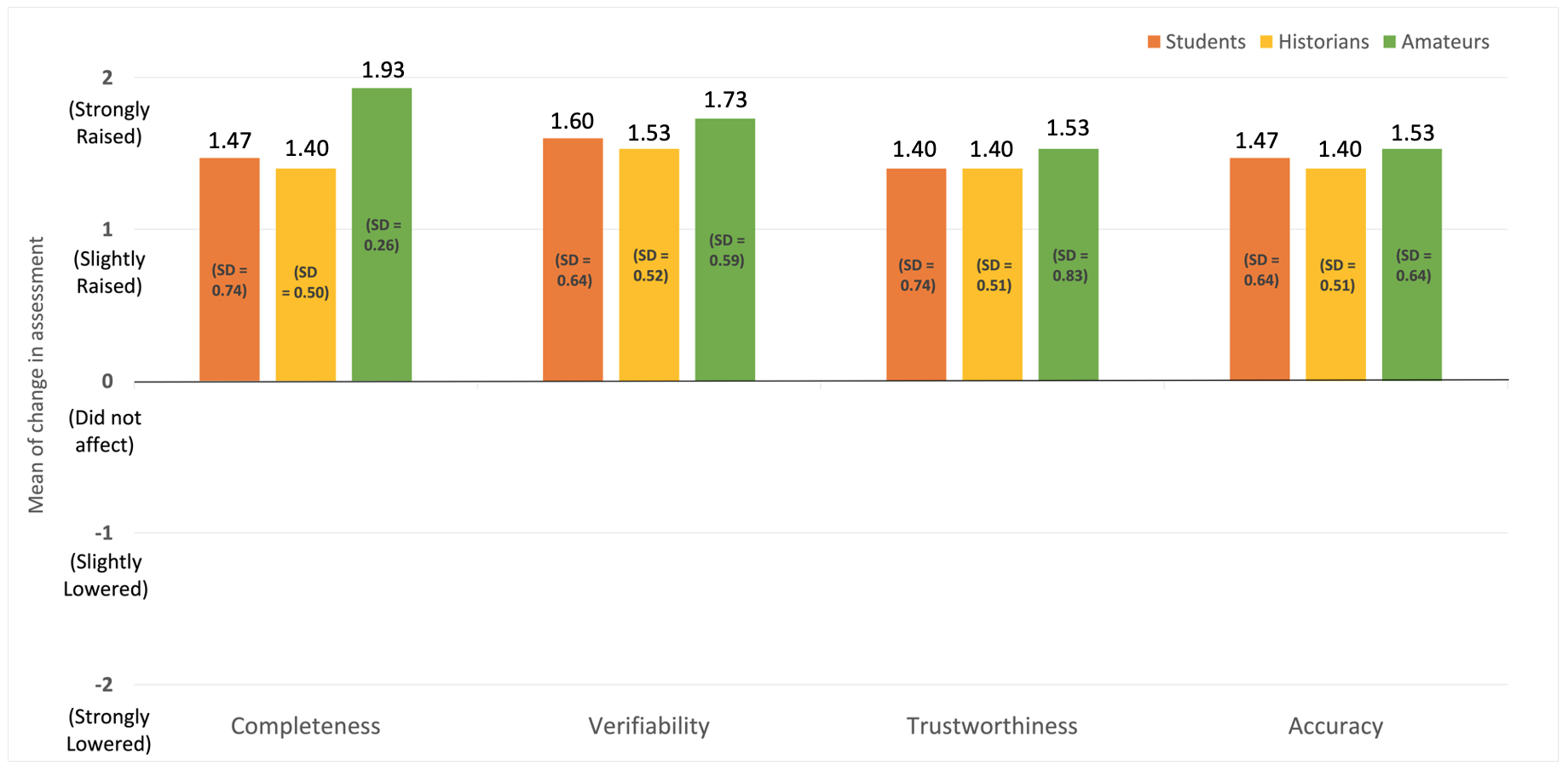}
    \caption{Summary of change in assessment of the information after using DoubleCheck.}
    \label{ref:assesstable}
\end{figure}

Overall, all 15 participants preferred the DoubleCheck version of CWPS over the original version for assessing (mean = 4.93 out of 5, SD = 0.25) the photo IDs. The participants first analyzed three photo IDs with the old interface, and assessed whether they were complete, accurate, trustworthy and verifiable. After analyzing the same photo IDs with the new interface, all participants reported that their assessment of information increased (see Figure~\ref{ref:assesstable}). 

For students, we see that the DoubleCheck version raised their assessment either strongly or slightly for almost all cases, barring a couple of cases where some students were sure of their assessment with the original interface and therefore, DoubleCheck did not affect their assessment. We did not see any scenario where DoubleCheck was responsible for lowering their assessment. 

Amateur experts reported that DoubleCheck raised their assessment strongly in most cases, and slightly for a few. There were a couple of cases, where the interface did not affect their assessment. In a few scenarios, we observed participants changing their opinion about the ID, or becoming more confident, after using DoubleCheck. 

Historians reported that the interface slightly raised their assessment in most scenarios, except a few scenarios where it strongly raised their assessment. We observed a couple of scenarios where the historians could not make a decision about the identity with the original interface, but became confident with the DoubleCheck interface. 

\subsection{Assessability Process}

Participants attributed multiple reasons for preferring the DoubleCheck version of CWPS. They found it to be more informative compared to the original system, and liked the way the information was organized and presented (mean = 4.80 out of 5, SD = 0.40). 
Some participants raised concerns about the information being overwhelming at first, but felt it was necessary for assessing the ID. In the words of H5, \emph{"It allows you to sort of digest the information at your own rate and choose how much you actually want to engage with the image. And I think that's just helpful in that it doesn't necessarily overwhelm you with too much information and unless you're actively making that choice to sort of bring it out.
"}

We observed a few cases where participants missed ID conflicts with the original interface, and assumed the ID displayed on top as the only proposed ID for the photo. We had to nudge them to notice the conflicting IDs. Participants pointed out that DoubleCheck’s layout was better equipped for handling such cases of ID conflicts. C3 said, \emph{"Here we are with this photo again, which in this case, it's coming up as Rishworth Rich, and has a verified ID, but it does have multiple identities. This is something that I mentioned earlier --- I wanted something right off the bat tells you that there's more than one ID on here."}

In addition to the layout and informative nature of DoubleCheck, participants mentioned the provenance information and stewardship visualizations (i.e., quality assessment badges), as compelling reasons for preferring the new system. In the following subsections, we report in detail our findings on how each of these DoubleCheck features helped participants in assessing the photo IDs.  

\subsubsection{Provenance}


\paragraph{\textbf{Users provided a wide range of identification sources as evidence for identifying a photo.}}

\begin{table}[]
\centering
\small
\begin{tabular}{|l|c|c|c|}
\hline
\multicolumn{1}{|c|}{\textbf{Source Type}} &
  \textbf{\# of IDs} &
  \textbf{\begin{tabular}[c]{@{}c@{}}\# of IDs with \\ Source Details\end{tabular}} &
  \textbf{\begin{tabular}[c]{@{}c@{}}\# of IDs with \\ Source Details \\ containing URLs\end{tabular}} \\ \hline
Period inscription without valediction                                                             & 1457                 & 1432                & 405                 \\ \hline
\begin{tabular}[c]{@{}l@{}}Period publication \\ (e.g. regimental history, newspaper)\end{tabular} & 1211                 & 1209                & 943                 \\ \hline
Other library, museum or archive                                                                   & 955                  & 950                 & 886                 \\ \hline
Period inscription with valediction                                                                & 425                  & 421                 & 98                  \\ \hline
Find A Grave                                                                                       & 291                  & 286                 & 278                 \\ \hline
Modern inscription                                                                                 & 146                  & 144                 & 94                  \\ \hline
\begin{tabular}[c]{@{}l@{}}Period documents  \\ (e.g., letter, album, diary, obituary, burial card, \\ pension records, discharge papers)\end{tabular} &
  75 &
  75 &
  7 \\ \hline
Family or descendant                                                                               & 68                   & 65                  & 1                   \\ \hline
\begin{tabular}[c]{@{}l@{}}Modern publication \\ (e.g. book, article)\end{tabular}                 & 45                   & 44                  & 27                  \\ \hline
\begin{tabular}[c]{@{}l@{}}Misc. websites / social media \\ (e.g. Facebook, Blogs)\end{tabular}    & 38                   & 37                  & 32                  \\ \hline
\textit{\textbf{Blank}}                                                                            & \textit{\textbf{37}} & \textit{\textbf{5}} & \textit{\textbf{1}} \\ \hline
Ancestry.com                                                                                       & 34                   & 31                  & \textit{10}         \\ \hline
Period inscription on album page                                                                   & 31                   & 28                  & 12                  \\ \hline
American Civil War Research Database (HDS)                                                         & 28                   & 26                  & 24                  \\ \hline
Library of Congress                                                                                & 27                   & 23                  & 22                  \\ \hline
US Army Heritage and Education Center (MOLLUS)                                                     & 21                   & 10                  & 9                   \\ \hline
Period inscription on Union case (containing photo)                                                & 21                   & 19                  & 13                  \\ \hline
Fold3                                                                                              & 18                   & 17                  & 1                   \\ \hline
Other genealogy website                                                                            & 15                   & 12                  & 8                   \\ \hline
National Archives                                                                                  & 12                   & 11                  & 4                   \\ \hline
Auction House Website                                                                              & 11                   & 11                  & 10                  \\ \hline
Other                                                                                              & 10                   & 9                   & 2                   \\ \hline
Dealer or collector                                                                                & 6                    & 4                   & 2                   \\ \hline
eBay listing                                                                                       & 6                    & 5                   & 5                   \\ \hline
\textbf{Total}                                                                                     & \textbf{4988}        & \textbf{4874}       & \textbf{2894}       \\ \hline
\end{tabular}%
\caption{Distribution of Sources provided for Pre-Identified Photos}
\label{tab:source_dist}
\end{table}

From our log analysis (see Table~\ref{tab:source_dist}), we found 4988 photo IDs on CWPS (DoubleCheck) where users had provided an original ID source (i.e., pre-identified photos). The most common source types were primary sources -- period inscriptions on the photo (1488) followed by other period sources, which included publications (1211) and documents (75). This was followed by websites of library, museum and archives (1014), including the U.S. Army Heritage and Education Center, Library of Congress, and National Archives. Users also cited other period sources, including documents (15) and publications (141). Apart from these primary and secondary scholarly sources, users also cited genealogy websites (138), including Ancestry.com and Find-a-Grave. Some others also cited word-of-mouth source types (74) such as a dealer/collector or descendant. 

Out of 4988 sources provided, only 114 were missing any details. Around 58\% of the source details contained a URL link, suggesting external evidence. We found 37 blank source types. 
    
\paragraph{\textbf{Users assessed the quality of a photo ID by verifying provenance information, and considered primary and secondary scholarly sources to be the most reliable.}}

In the lab study, participants first assessed the photo IDs using the original CWPS system, which only had the photo and biography source. A large part of their assessment process focused on verifying the source information available in the original interface and whether they considered it reliable or not. We observed that many of our participants did not notice the available source information in the old interface at first. 
    
All participants considered period inscriptions with the person’s name on the photo as gold-standard proof for the photo’s ID. C4 expressed confidence seeing a period inscription: \emph{"I think this one also has his name on it. That looks to be a period identification. I don't have any doubt that that is Joseph B. Carr from what I'm seeing here."} At the same time, C3 raises concerns that the period inscription could be a modern forgery: \emph{"The one concern I have with this inscription is how shaky it is and how it bleeds into the paper, which sends up a couple of red flags for me."}
    
They also rated scholarly sources, such as libraries, museums and archives, as highly trustworthy sources, compared to other online databases. H2, after seeing a source that he did not recognize, said, \emph{"There's no identifying source like the Library of Congress. For example, the other one had that listed as a resource and I tend to trust the Library of Congress. The other soldier database, not really familiar with it. So I'm not sure if that's a trustworthy site or not."}
    
Participants had mixed opinions about personal collections being cited as sources. Some found them useful, as the owner may have done extensive research. In C3’s words, \emph{"For the most part, these artifact collectors of Civil War photography, they spend a lot of money investing into these. They want to make sure that what they're buying is authentic. [...]
I do hold people who upload images from a personal collection to a pretty high standard compared to somebody who's uploaded something from like Ancestry or Find a Grave, for example."}

However, participants generally found it difficult to verify someone’s word-of-mouth as evidence for a photo’s ID. H2 said, \emph{"The way these stories go, you know somebody has purchased a CDV [carte de visite] at an antique store as they were driving through a town or it may have been in the family for a long time, and the words have passed down. So it's hard to verify right if it's just based on that information alone."}
    
With the original system, participants also experienced difficulty in immediately verifying sources that they were not familiar with, especially when they lacked an obvious institutional affiliation. C1, while assessing an ID conflict where one ID had a period inscription and the other one came from a public library, said: \emph{"This is a public library which is fine, but I trust the Maine State Archives just because I've done so much work with their stuff. Bangor Public Library, I don't know too much about."} 
For example, S1 had concerns about a website that he did not recognize: \emph{"It makes me want to question the validity of it just because it's not from a university or museum like a .gov or .org. It's simply a website that somebody put together.
"}
    
Historians also expressed a desire to see richer provenance information, direct links to primary sources, or multiple corroborating sources about the photo’s identity while using the original system. H3 said, \emph{"The (American Civil War) research database, that's helpful for sure, but then having the direct source also available is more trustworthy to me just seeing like the original source, I'd want to see that, though I think the link is definitely helpful as a secondary source. So maybe a series of sources, if there are multiple places where people are acquiring this from, whether it's the database or the book, that might be helpful."} 
    
\paragraph{\textbf{Users found DoubleCheck’s provenance visualization to be informative and useful for verifying photo IDs.}}

While assessing the photo IDs with the original CWPS system, participants largely focused on verifying the reliability of the available source information (i.e., photo source and biography source). However, we observed that many of the participants did not notice the source information at first. All participants rated period inscriptions with the person’s name on the photo and scholarly sources, such as libraries, museums and archives, as highly trustworthy sources. At the same time, participants experienced difficulty in verifying sources that they were not familiar with, including someone’s word-of-mouth as evidence for a photo ID. 

With DoubleCheck, participants were easily able to access the ID sources section, where the different sources linked to a given ID were listed. They found the provenance information easy to understand (mean = 4.73 out of 5, SD = 0.44) and useful for assessing the ID (mean = 4.80 out of 5, SD = 0.40). We observed active users scrolling down directly to the ID sources section to investigate the sources. 
C3, said, \emph{"Finding those sources I think I can do a lot quicker and much more easily on this newer system compared to the old one."}

All participants liked the way the sources were organized into different categories (primary, scholarly or non-scholarly secondary, etc.) and found that helpful for assessing the source credibility. S1 said, \emph{"I think it's helpful because it gives me an idea where the sources are from, if it’s scholarly or not, because you know the sources are very important when doing historical research."} C3 summed it up: \emph{"I find it very clear and as I'm scrolling through these, it has these title bars here that really stand out to me and break it down. The primary sources --- it is at the top, the most reliable one, and then you get into the lower categories as you go down. The way it's broken out, it's very clear.
"}
    
Participants not only liked the high-level categorization of the sources, but also how the sources were presented (i.e., source types and details, along with the photo). S3 appreciated the sources being categorized and visually identified with emoji icons, explaining, \emph{"This one was really helpful. [...]
Here, I know what type of source before I even click it on. I like that because a lot of times, the type of source is pretty important. If it’s like something from a book, then the information needs to be published and verified."} S2, on seeing a `dealer/collector' source type, analyzed the ID more critically: \emph{"
I think that it's appropriate to put this here because I think seeing these words --- ‘dealer or collector’ earlier on the page would cast doubt on other things." }
    
Participants found the level of detail in the presentation of the provenance information to be appropriate for their assessment. H3 said, \emph{"I liked the level of detail in terms of how the sources were listed, where things were found, whether or not the ideas have been verified. Having that right at the top was a great little signifier for me. I think that affirmed my opinions, for sure." } Similar sentiments were echoed by H4: \emph{"
I want to see the secondary sources that backed this up honestly, the primary sources that back this up. I think this is really helpful because you get this very careful step-by-step of why does this support, why is this a source at all? [...]
It's much more convincing.
"}


\subsubsection{Quality Assessment Badges}

\paragraph{\textbf{Users found the system’s quality assessment badges and visualization to be helpful for making initial assessments about a photo’s ID.}}

\begin{table}[]
\small
\centering
\begin{tabular}{|c|c|c|}
\hline
\textbf{Quality Badge} &
  \textbf{\begin{tabular}[c]{@{}c@{}}\# of IDs\\ (Pre-Identified \\ Photos)\end{tabular}} &
  \textbf{\begin{tabular}[c]{@{}c@{}}\# of IDs\\ (Post-Identified \\ Photos)\end{tabular}} \\ \hline
\textbf{Needs Tags}         & 58   & 3   \\ \hline
\textbf{Needs Verification} & 3012 & 152 \\ \hline
\textbf{Verified ID}        & 1918 & 13  \\ \hline
\end{tabular}
\caption{Distribution of Quality Badges}
\label{tab:quality_badge_dist}
\end{table}

\begin{table}[]
\small
\centering
\begin{tabular}{|l|c|}
\hline
\multicolumn{1}{|c|}{\textbf{Source Type}} &
  \textbf{\# of IDs} \\ \hline
Family or descendant                           & 59            \\ \hline
Find A Grave                                   & 290           \\ \hline
\begin{tabular}[c]{@{}l@{}}Misc. Websites / social media \\ (e.g. Facebook, Blogs)\end{tabular} &
  37 \\ \hline
Ancestry.com                                   & 32            \\ \hline
\textit{Blank}                                 & \textit{26}   \\ \hline
Period inscription without valediction         & 16            \\ \hline
US Army Heritage and Education Center (MOLLUS) & 21            \\ \hline
Fold3                                          & 18            \\ \hline
Other library, museum or archive               & 945           \\ \hline
\begin{tabular}[c]{@{}l@{}}Period publication \\ (e.g. regimental history, newspaper)\end{tabular} &
  1200 \\ \hline
Modern inscription                             & 142           \\ \hline
\begin{tabular}[c]{@{}l@{}}Period documents  \\ (e.g., letter, album, diary, obituary, burial card, \\ pension records, discharge papers)\end{tabular} &
  71 \\ \hline
Period inscription on album page               & 1             \\ \hline
Other genealogy website                        & 15            \\ \hline
Modern publication (e.g. book, article)        & 44            \\ \hline
Library of Congress                            & 25            \\ \hline
Dealer or collector                            & 5             \\ \hline
National Archives                              & 12            \\ \hline
American Civil War Research Database (HDS)     & 26            \\ \hline
Auction House Website                          & 11            \\ \hline
Other                                          & 8             \\ \hline
Period inscription with valediction            & 3             \\ \hline
eBay listing                                   & 5             \\ \hline
\textbf{Total}                                 & \textbf{3012} \\ \hline
\end{tabular}
\caption{Source distribution for pre-identified photos that "Need Verification"}
\label{tab:source_dist_nv}
\end{table}

From our logs (see Table~\ref{tab:quality_badge_dist}), we found 1918 pre-identified photos and 13 post-identified photos had a "Verified ID" badge. Upon further investigation, we found that 1902 of these pre-identified photos were verified due to reliable provenance information (i.e., primary sources). Of the remaining 16, only 2 of these IDs were verified by community consensus over a scholarly secondary source. The rest (13) were found to be replicas of other verified IDs (with primary sources). Similarly, 12 out of the verified 13 post-identified photos were found to be replicas of other verified IDs (with primary sources). Only one photo was verified by community consensus over a facial match with another verified ID. Majority of the IDs (61.3\%) had a "Needs Verification" badge. 

From Table ~\ref{tab:source_dist_nv}, we find that a significant number of IDs with scholarly secondary sources such as period publications, libraries, museums, and archives remain unverified due to limited community participation in verifying them. Further, we also found that a few IDs with period inscriptions were unverified due to the presence of a conflicting ID (mostly duplicates). 

Participants liked that the quality assessment badges were prominently displayed and were easy to understand (mean = 4.67 out of 5). 
The "Needs Verification" badge made participants want to critically analyze the photo ID for themselves. H2 compared it to the content assessment ratings on Wikipedia: \emph{"
I'm not saying that it's definitely wrong, but there needs to be more information to prove that it's right or to give additional, you know, additional sources that I can click and check for myself and see if it's right." }
H5 agreed, \emph{"I think immediately seeing that `Needs Verification' quite clearly and so that affirms to me that this is something in contention and that I should be cautious in looking at it."}
    
    
    
The "Verified ID" badge, on the other hand, made users feel more confident about the photo ID being accurate. C5 believed it would save him a lot of time: \emph{"[R]ight now with this verified ID, I'm going to trust it without having to do too much looking into it." }
H3 felt confident seeing the ID verified by the system: \emph{"It gives me the sense that, okay, the information is likely correct if it's been verified by several different sources and visitors."} S4 compared it to the blue verification checkmark on Twitter. 
    
    
Participants, especially historians, found the checklist visualization effective at conveying the current quality assessment status of the ID. H1 found it to be "very transparent". 
H4 explained how she approached the checklist as a historian: \emph{"My first thought would be to scroll down and look at where in the bubble timeline it was, and then I would see it as this challenge of like, 'Okay, well, can I help?' Would I use this as a source without doing more digging? No. I think it needs verification. Whatever this says is whatever historians are going to do, right?"} 

Student participants appreciated the quality badges as a set of guidelines to refer to and augment their developing knowledge. S1 said, \emph{"It makes you feel a little bit better that there’s some kind of vetting process."} S2 said that the process was \emph{"scholarly and professional"} and found it to be \emph{"definitely effective." }
    
Some participants, especially amateur experts like C2 and C5, did not read too much into the badges. C5 said, \emph{"Maybe it doesn't sound right, but I'm always more critical in looking at these things. I try to be as precise as possible. I'm basically always skeptical when I go into it."}
    
Overall, the participants found the quality assessment badges to be useful for assessing the ID (mean = 4.80 out of 5, SD = 0.40). 

\paragraph{\textbf{Users approved DoubleCheck’s vetting process for determining the quality badges.}}
We observed that DoubleCheck’s quality assessment lined up with the participant’s own assessment in most cases. Whenever it did not match, the participants analyzed all available information (community opinions and the sources) and eventually changed their assessment to the correct one. 

Some participants asked for details of how the quality assessment process works, and were satisfied with the multi-fold checks and balances involved. H1, after understanding how it works, said, \emph{"Yeah, I do feel more confident in your (system’s) assessment. I do feel willing to accept your assessment and I would not feel the need to go through Sheridan and look at all the details in the same way that I would need to with Brown and Cook." }

H4, while assessing an ID that needed verification, pointed out that the quality badge should also be displayed next to the biography and service records as \emph{"you lose that uncertainty when you look at this part of the page."}


 \section{Discussion}

\subsection{Encouraging Accurate Provenance for Preventing Misinformation}

To help users assess the credibility of sources, DoubleCheck displays all available ID sources for a given, organized by different high-level categories, drawing parallels to the provenance visualizations proposed by Forte et al. \cite{forte2014designing} for Wikipedia. However, in the Wikipedia context, the visualization assumes the prior existence of the data (i.e., source information and categories). Here, we built an architecture, as part of DoubleCheck, to capture accurate provenance information in CWPS, which the community used for uploading a wide range of sources for over 4900 IDs. CWPS also supports the complex creative endeavor of original research, in contrast to Wikipedia. We engaged with the community and facial recognition to indicate whether a photo source, that is linked via facial similarity, can be reliably used as evidence for the ID or not. 

DoubleCheck's provenance visualization allowed users to easily discover and process the provenance information if it is credible. Displaying the source types was particularly useful for novices, who were not as familiar with the different sources as the experts, as the knowledge of whether the source is a book, library or museum, blog post, or someone’s personal collection can help in quick assessments.

Many photos are identified on the basis of word-of-mouth, be it from a dealer or a collector or a claimed descendant, and users found it difficult to verify such IDs. However, they also felt it was important to highlight that information to avoid being misled. Our provenance visualizations support personal collections through dedicated source types for word-of-mouth identifications ("Dealer or Collector," "Descendant") and display them prominently. Further, they are grouped under the non-scholarly secondary category, which are considered least trustworthy by the quality assessment engine and are assigned a "Needs Verification" badge. Our findings showed the provenance visualization, an outcome of Design Goal 1, to be highly informative and effective for users in assessing different photo IDs, thus supporting prior work on assessability frameworks \cite{forte2014designing,wiggins2016community}. 

CWPS, unlike Wikipedia, supports original research be it for verifying whether two photos show the same person or not, or for establishing the ID of a photo. When two photos are matched to each other (i.e., show the same person), they essentially become corroborating evidence. In addition to the original sources, DoubleCheck also supports photo sources that are linked via facial similarity.

The system, however, relies on users to manually verify the source information whether it says that the photo is the claimed person or not, who can then leave their vote of confidence and justification for other users through the validation workflow. We observed the effectiveness of this process in our study. However, to have a better understanding of the accuracy of the source information on CWPS, we aim to conduct a large-scale study of CWPS ID sources as future work. Drawing inspiration from prior work on automated fact-checking \cite{cazalens2018content,zhang2021your}, we are also exploring crowd-AI workflows to verify whether the information actually comes from the source mentioned, and whether the source is credible or not. By engaging with relevant domain knowledge for determining different source types and their trustworthiness, DoubleCheck’s provenance architecture and visualizations can generalize to other person identification tasks~\cite{reese2022emerging}.

\subsection{Assessing Photo Identifications with the help of Quality Indicators}

We built a quality assessment engine to automatically verify IDs based on provenance and community opinions. Over 1900 IDs were found to be verified on CWPS. Users, upon receiving an explanation of the vetting process, approved of the checks and balances involved in the process. We also designed badges and visualizations to show the current quality status of the ID. We found the quality assessment badges, a result of Design Goal 2, to be largely helpful for users in making quick assessment of the IDs. 

The "Needs Verification" badges were effective in making the users cautious about the ID, which encouraged them to critically analyze it for themselves. On the other hand, the "Verified ID" badges made users feel confident about the accuracy of the ID. This was particularly useful for novices, who were more willing to trust the system’s assessment, as they might be prone to making incorrect assessments themselves. \vm{Further, we found that the "Verified ID" badges helped the users resolve identity conflicts when multiple identities were proposed (see Figure~\ref{fig:allids}).} Our efforts towards being transparent about ID quality on CWPS draw parallels to recent work on the promising effects of misinformation labels on social media content \cite{nassetta2020state}. 

\vm{The misinformation-countering potentials of the "Verified ID" badge open up possibilities for DoubleCheck to become a database repository of Civil War photo verifications beyond CWPS. Civil War photos are sold and auctioned in large quantities on numerous online platforms including eBay, auction house websites, Facebook Marketplace, etc.~\cite{ccwp_2023}. However, unlike other collectibles, such as baseball card authentication~\cite{jamal2011mandated}, there does not exist any universally-accepted authentication service for Civil War photos. Thus, Civil War photo transactions are exposed to risks, such as the spread of misinformation or the devaluation of a photograph, caused by misidentifications. This situation presents opportunities for creating augmentation tools, such as DoubleCheck-powered browser plugins, that can help authenticate Civil War photos on these marketplaces. Engaging more community participation and expanding the database to include more "Verified IDs" could help establish DoubleCheck as a trustworthy resource for historians, dealers, collectors, curators, and others interested in Civil War-era photography.}

While showing the quality badges for the ID was our starting point, extending verification to other sources on the page, such as biographical information and service records, can help in correctly shaping the user’s assessment, as H4 pointed out.  One of the active users, C1, raised the issue of certain IDs from reputable scholarly sources being in the "Needs Verification" stage, as they required consensus from the community. While these IDs can eventually be verified through more active participation, we believe this precautionary measure of seeking additional verification from the community is necessary even for scholarly sources. There are numerous instances where IDs made by scholarly sources have been disproven \cite{luther2020real,luther2019gold}, including one of the photos used in this study. 

A possible design implication is to add an intermediate category for such IDs (coming from scholarly sources) that \vm{are \emph{easily verifiable} and could require a lower voting threshold.} This would separate such \emph{easily verifiable} IDs from the other IDs that have been identified on the basis of biographical information and facial similarity to another photo, which require more eyes. Another possible addition to the badges could be to distinguish IDs that are \emph{unverifiable} until new sources are provided, which would include low-trust sources such as personal collections, genealogy websites, etc. \vm{We found a large proportion of IDs that "need verification" (see Table~\ref{tab:quality_badge_dist}), so introducing these new granular conditions might help CWPS move towards a more representative distribution.}

The flexibility of DoubleCheck's framework allows it to incorporate source trustworthiness information according to the domain in question, and as a result, can be extended for other types of historical person identification beyond Civil War photos. For example, open-source intelligence (OSINT) investigators, who analyze publicly available data like social media for investigating matters of national security, law enforcement, and journalism, often deal with person identification tasks in a modern context \cite{wong2016fluidity,roberts2021importance}. Modifying DoubleCheck's stewardship rules for consensus also allows it to be adapted for the needs of the OSINT community, which are often high-stakes in nature and might require more stringent measures. Recent efforts towards crowdsourced fact-checking on social media such as Twitter's Birdwatch~\cite{noauthor_introducing_2021} can also benefit from DoubleCheck's approach by incorporating trustworthiness information for known sources and displaying stewardship visualizations.

\section{Conclusion}

DoubleCheck attempts to help users assess photo IDs better on CWPS. We present a quality assessment framework that builds on the concepts of information provenance and stewardship, and adapts it for the task of historical person identification. We demonstrate the effectiveness of DoubleCheck on CWPS, an existing online platform, where users found the provenance information and stewardship visualizations (i.e., the quality assessment badges) useful for making accurate assessments of photo IDs on the platform. This work opens doors for exploring new ways to engage the community and AI for building assessable online information systems for historical photo identification.

\begin{acks}
\vm{This research was supported by NSF IIS-1651969 and a Virginia Tech ICTAS Junior Faculty Award.}
\end{acks}

\bibliographystyle{ACM-Reference-Format}
\bibliography{sample-base}

\clearpage
\appendix


\section{Survey Questionnaire}

\begin{table}[H]
\setlength{\tabcolsep}{5pt} 
\renewcommand{\arraystretch}{1.5} 
\begin{tabular}{|l|l|l|}
\hline
\multicolumn{1}{|c|}{\textbf{\begin{tabular}[c]{@{}c@{}}Interface \\ Condition\end{tabular}}} &
  \multicolumn{1}{c|}{\textbf{Questions}} &
  \multicolumn{1}{c|}{\textbf{Responses}} \\ \hline
\multirow{4}{*}{\begin{tabular}[c]{@{}l@{}}Old Interface \\ (CWPS)\end{tabular}} &
  The information available on the page is \textit{complete}. &
  \multirow{4}{*}{\begin{tabular}[c]{@{}l@{}}1. Strongly Disagree\\ 2. Slightly Disagree\\ 3. Slightly Agree\\ 4. Strongly Agree\end{tabular}} \\ \cline{2-2}
 &
  The information available on the page is \textit{verifiable}. &
   \\ \cline{2-2}
 &
  The information available on the page is \textit{trustworthy}. &
   \\ \cline{2-2}
 &
  The information available on the page is \textit{accurate}. &
   \\ \hline
\multirow{4}{*}{\begin{tabular}[c]{@{}l@{}}New Interface \\ (CWPS + DoubleCheck)\end{tabular}} &
  \begin{tabular}[c]{@{}l@{}}This interface \_\_\_\_\_\_\_\_\_\_ my assessment about the information\\ on the page being \textit{complete}.\end{tabular} &
  \multirow{4}{*}{\begin{tabular}[c]{@{}l@{}}1. Strongly Lowered\\ 2. Slightly Lowered\\ 3. Did not affect\\ 4. Slightly Raised\\ 5. Strongly Raised\end{tabular}} \\ \cline{2-2}
 &
  \begin{tabular}[c]{@{}l@{}}This interface \_\_\_\_\_\_\_\_\_\_ my assessment about the information\\ on the page being \textit{verifiable}.\end{tabular} &
   \\ \cline{2-2}
 &
  \begin{tabular}[c]{@{}l@{}}This interface \_\_\_\_\_\_\_\_\_\_ my assessment about the information\\ on the page being \textit{trustworthy}.\end{tabular} &
   \\ \cline{2-2}
 &
  \begin{tabular}[c]{@{}l@{}}This interface \_\_\_\_\_\_\_\_\_\_ my assessment about the information\\ on the page being \textit{accurate}.\end{tabular} &
   \\ \hline
\end{tabular}
\caption{Comparative survey questions answered by the user for each of the three photos during the lab study.}
\label{tab:comparative questions}
\end{table}

\section{Stewardship Visualizations}

\begin{figure}[htbp!]
\centering
\begin{subfigure}{.50\textwidth}
  \centering
  \includegraphics[scale=0.12]{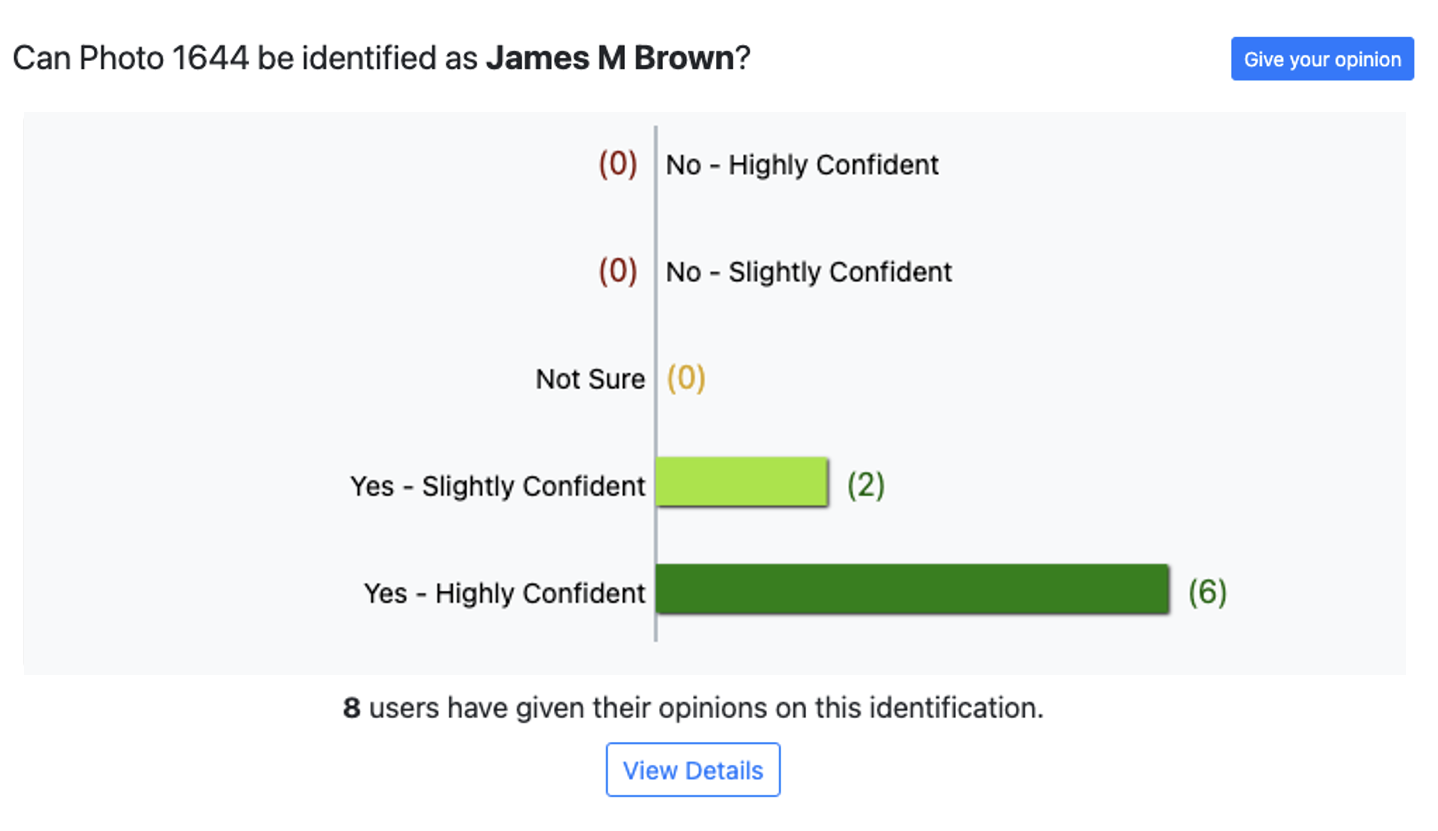}
  \caption{Community's confidence on the ID}
  \label{fig:idchart}
\end{subfigure}%
\begin{subfigure}{.50\textwidth}
  \centering
  \includegraphics[scale=0.17]{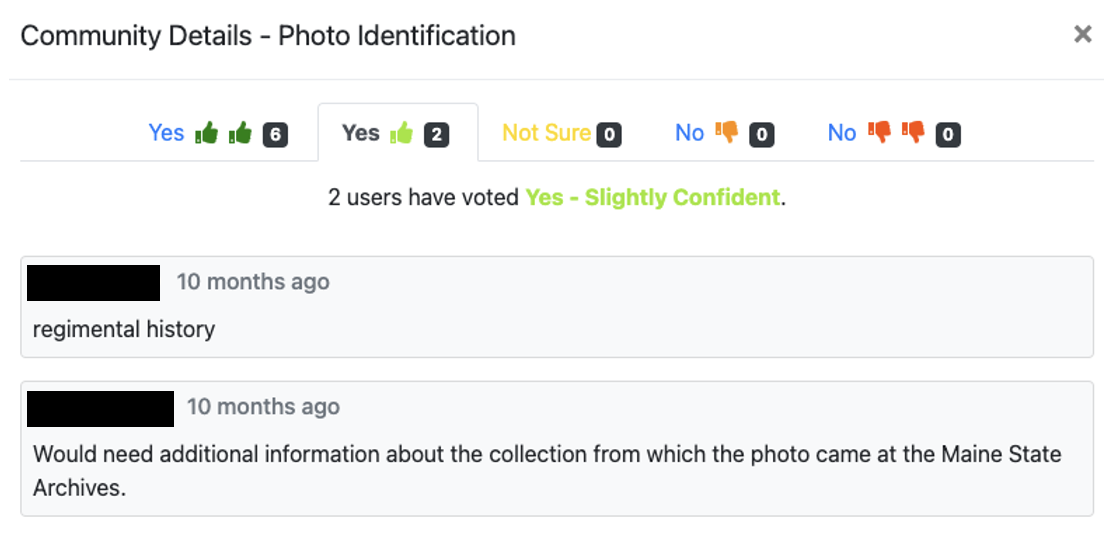}
  \caption{Community's notes justifying their votes.}
  \label{fig:viewdetails}
\end{subfigure}
\caption{ID Reliability Visualization}
\label{fig:idreliabilityvisalizations}
\end{figure}

\section{Validation Interface}

\begin{figure}[H]
    \centering
    \includegraphics[scale=0.20]{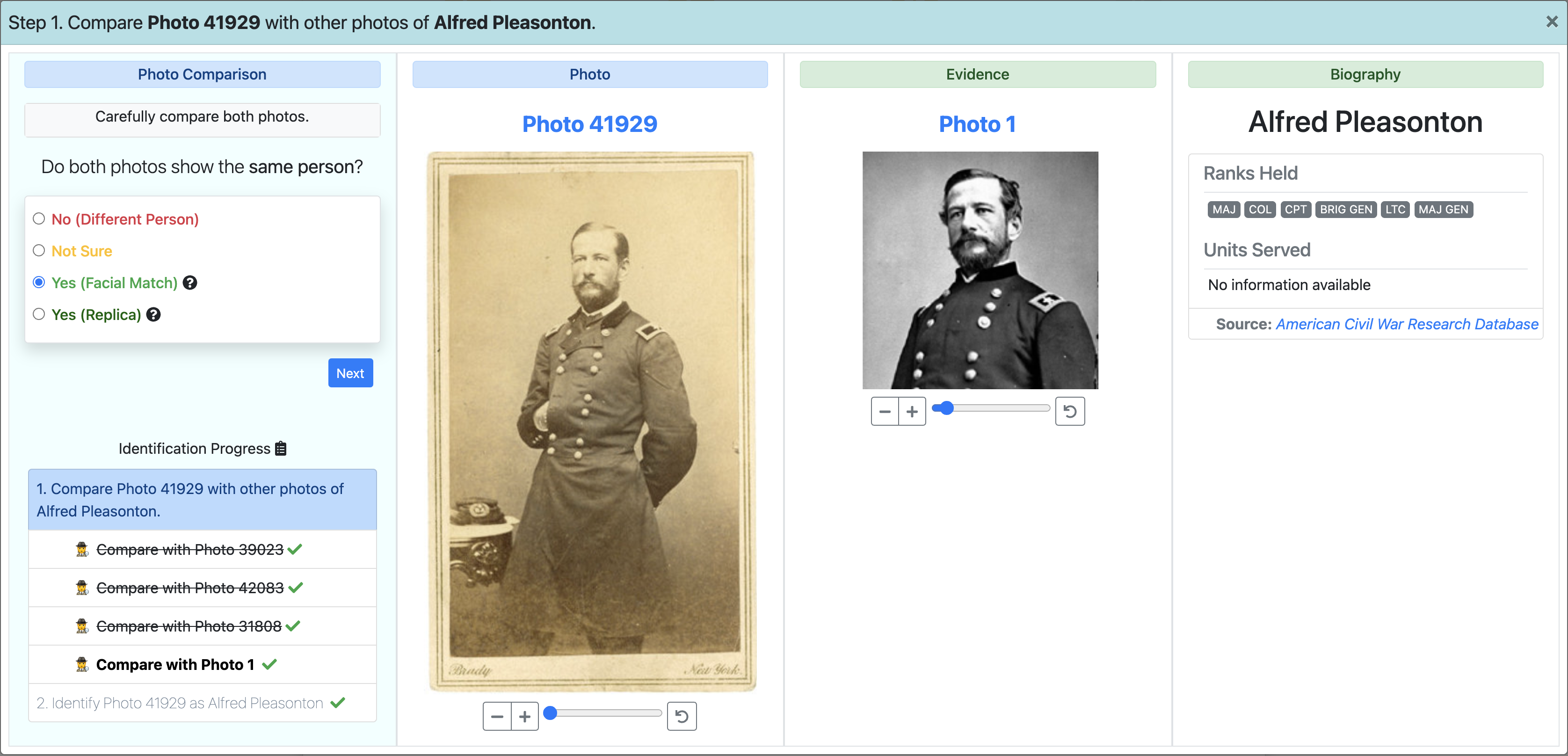}
    \Description{Validation Interface (Step 1 of Photo Steward's Validation Workflow). Users can compare two photos and answer whether they show the same person or not. The query photo is shown in the Photo column (second from left) and the target photo is shown under the Evidence column (third from left). In the first column from the left --- "Photo Comparison", they have the option of selecting whether the two photos are a facial match (i.e., same person, different views) or a replica (i.e., same person, same view). Here, the user is comparing whether Photo 41929 and Photo 1 show the same person or not.}
    \caption{Validation Interface (Step 1 of Photo Steward's Validation Workflow). Users can compare two photos and answer whether they show the same person or not. They have the option of selecting whether the two photos are a facial match (i.e., same person, different views) or a replica (i.e., same person, sane view). Here, the user is comparing whether Photo 41929 and Photo 1 show the same person or not. If multiple faces are available for the same ID, they appear one after the other in the order in which they were uploaded to CWPS.}
    \label{fig:comparisonstep1}
\end{figure}

\begin{figure}[H]
    \centering
    \includegraphics[scale=0.25]{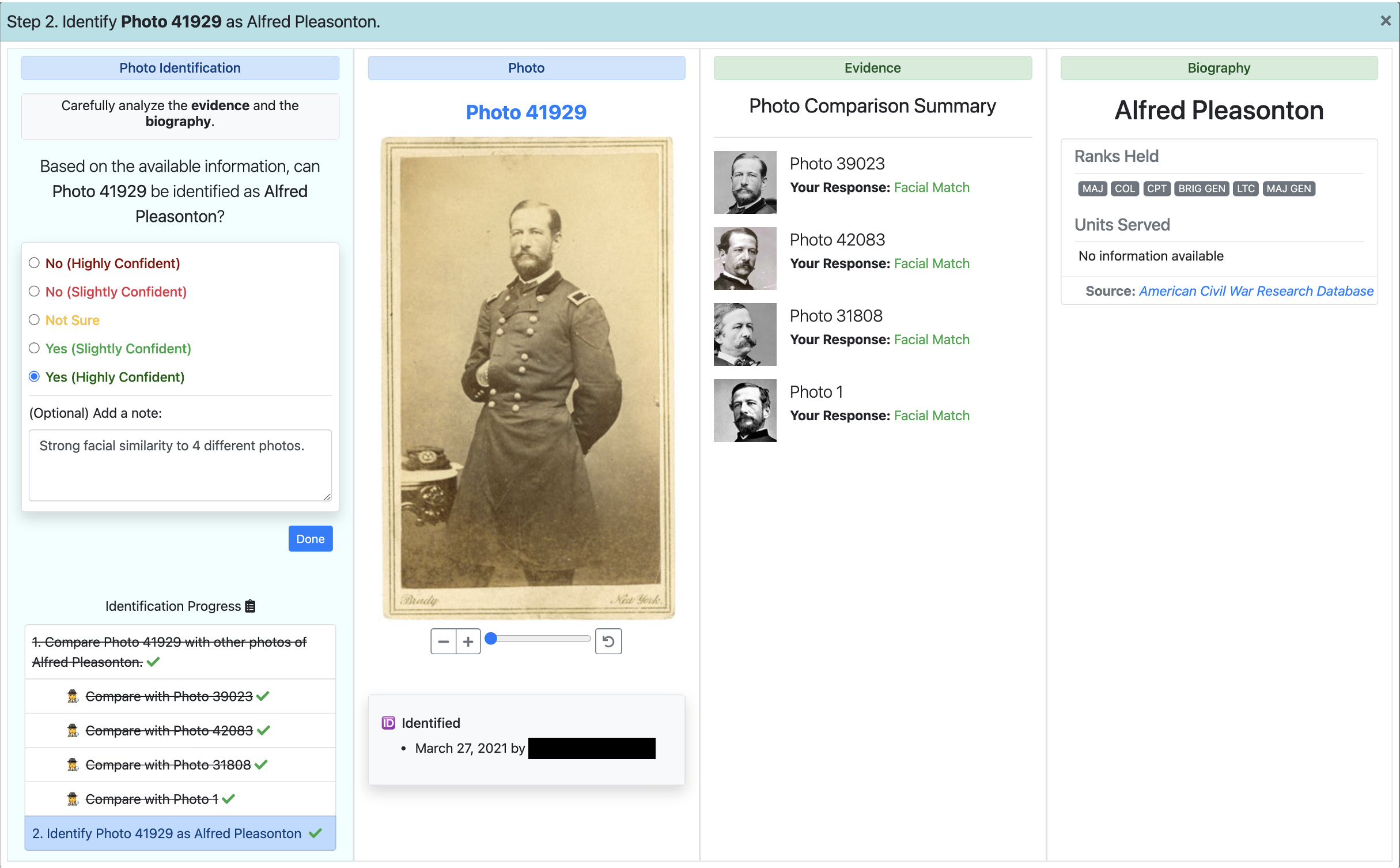}
    \Description{Validation Interface (Step 2 of Photo Steward's Validation Workflow). Users vote on whether the query photo (shown in the Photo Column) can be identified as the target identity (shown in the Biography column -- right most column) by expressing their confidence (in the left most column that says "Photo Identification"). They can also add an optional note to justify their decision.}
    \caption{Validation Interface (Step 2 of Photo Steward's Validation Workflow). Users vote on whether the query photo can be identified as the target identity by expressing their confidence. They can also add an optional note to justify their decision. The evidence panel displays a summary of the user's responses from the first step, where the faces are displayed next to the user's facial similarity comparison with the query photo. The faces are ordered in the way they appear for comparison, i.e., the order in which they were uploaded to CWPS.}
    \label{fig:comparisonstep2}
\end{figure}

\end{document}